\documentclass[useAMS,usenatbib]{mn2e}
\pdfoutput=1
\usepackage[T1]{fontenc}
\usepackage[colorlinks,linkcolor=red,citecolor=blue,urlcolor=blue ]{hyperref}
\usepackage{graphicx}
\usepackage{amsmath,amssymb}

\newcommand{\mhimh}{M_{\rm HI}(M_h)}
\newcommand{\mhi}{M_{\rm HI}}
\newcommand{\Mpc}{\, h^{-1} \, {\rm Mpc}}

\title[The HI content of dark matter halos at $z\approx 0$ from ALFALFA]{The HI content of dark matter halos at $z\approx 0$ from ALFALFA}
\author[A. Obuljen et al.]{Andrej Obuljen$^{1,2}$\thanks{aobuljen@sissa.it}, David Alonso$^{3,4}$, Francisco Villaescusa-Navarro$^5$, Ilsang Yoon$^6$, \newauthor{Michael Jones$^7$}\\
  $^1$SISSA- International School for Advanced Studies, Via Bonomea 265, 34136 Trieste, Italy\\
  $^2$INFN -- National Institute for Nuclear Physics, Via Valerio 2, I-34127 Trieste, Italy\\
  $^3$School of Physics and Astronomy, Cardiff University, The Parade, Cardiff, CF24 3AA, UK\\
  $^4$University of Oxford, Denys Wilkinson Building, Keble Road, Oxford OX1 3RH, UK\\
  $^5$Center for Computational Astrophysics, 162 5th Ave, New York, NY, 10010, USA\\
  $^6$National Radio Astronomy Observatory, 520 Edgemont Road, Charlottesville, VA, 22903, USA\\
  $^7$Instituto de Astrof{\'i}sica de Andaluc{\'i}a, Glorieta de la Astronom{\'i}a, Granada, 18008, Spain}

\begin{document}
  \date{\today}
  \pagerange{1--17} \pubyear{2018}
  \maketitle

  \begin{abstract}
    We combine information from the clustering of HI galaxies in the 100\% data release of the Arecibo Legacy Fast ALFA survey (ALFALFA), and from the HI content of optically-selected galaxy groups found in the Sloan Digital Sky Survey (SDSS) to constrain the relation between halo mass $M_h$ and its average total HI mass content $\mhi$. We model the abundance and clustering of neutral hydrogen through a halo-model-based approach, parametrizing the $\mhimh$ relation as a power law with an exponential mass cutoff. To break the degeneracy between the amplitude and low-mass cutoff of the $\mhimh$ relation, we also include a recent measurement of the cosmic HI abundance from the $\alpha$.100 sample. We find that all datasets are consistent with a power-law index $\alpha=0.44\pm 0.08$ and a cutoff halo mass $\log_{10}M_{\rm min}/(h^{-1}M_\odot)=11.27^{+0.24}_{-0.30}$. We compare these results with predictions from state-of-the-art magneto-hydrodynamical simulations, and find both to be in good qualitative agreement, although the data favours a significantly larger cutoff mass that is consistent with the higher cosmic HI abundance found in simulations. Both data and simulations seem to predict a similar value for the HI bias ($b_{\rm HI}=0.875\pm0.022$) and shot-noise power ($P_{\rm SN}=92^{+20}_{-18}\,[h^{-1}{\rm Mpc}]^3$) at redshift $z=0$.
  \end{abstract}
  

  \section{Introduction}\label{sec:intro}

    The $\Lambda$CDM model has become the most successful theory framework that is able to explain a wide variety of cosmological observations, from the temperature and polarization anisotropies in the cosmic microwave background (CMB) \citep{2016A&A...594A..13P} to the spatial distribution of galaxies at low redshift \citep{2017MNRAS.470.2617A}. Some of the free parameters of this model are connected with open questions in fundamental physics, such as possible deviations from a pure cosmological constant, the cosmic abundance of dark matter or the sum of neutrino masses. The main aim of modern cosmological experiments is to determine the value of those parameters with the best possible combination of precision and accuracy.

    In this endeavour, the statistics of the matter distribution contains an enormous amount of information to potentially constrain the value of these cosmological parameters. Unfortunately, the matter distribution is not directly observable, but can only be inferred through tracers of it such as galaxies, quasars and cosmic neutral hydrogen. In particular, 21cm intensity mapping \citep{Battye:2004re,2006ApJ...653..815M,Chang:2007xk,2008MNRAS.383..606W,2008PhRvL.100p1301L,2009astro2010S.234P,Bagla:2009jy,2013MNRAS.434.1239B,2013ApJ...763L..20M,2013MNRAS.434L..46S,2015ApJ...803...21B,2018MNRAS.tmp..340A} has recently become one of the main contenders in the quest to map out the three-dimensional cosmic density field out to the highest possible redshifts. In spite of the significant observational challenges of this technique, mostly associated with the presence of strong and complex radio foregrounds \citep{2005ApJ...625..575S,2014MNRAS.441.3271W,2015PhRvD..91h3514S,2015MNRAS.447..400A,2015arXiv151005453W}, intensity mapping (IM) offers a unique way to produce fast and economical, three-dimensional maps of the overdensity of neutral hydrogen (HI) in the Universe. For this reason, intensity mapping has been put forward as an ideal method to probe cosmology on large scales.

    However, the properties of HI, especially in terms of clustering, are still not fully understood. This is due to a number of reasons: the early stage of IM as an observational probe, the difficulty of detecting the faint 21cm line for a sufficiently large number of sources at high redshifts, and the possibly conflicting evidence \citep{2017MNRAS.471.1788C} coming from observations of low-redshift HI surveys \citep{2005MNRAS.359L..30Z,2012ApJ...750...38M}, the Lyman-$\alpha$ forest \citep{2005MNRAS.364.1467Z,2012A&A...547L...1N,2013A&A...556A.141Z,Crighton} and the clustering of damped Lyman-$\alpha$ systems \citep{2006ApJ...636..610R,2018MNRAS.473.3019P}. Understanding HI is vital both for cosmology and astrophysics, since it also plays a vital role in understanding the star formation history \citep{1998ApJ...498..541K}.
    
    At linear order, the amplitude of the 21cm power spectrum at redshift $z$ is proportional to the product of the HI bias $b_{\rm HI}(z)$ and its cosmic abundance $\Omega_{\rm HI}(z)=\rho_{\rm HI}(z)/\rho_c(z=0)$, where $\rho_{\rm HI}(z)$ is the mean HI density at redshift $z$ and $\rho_c(z=0)$ is the critical density at $z=0$. While the value of $\Omega_{\rm HI}(z)$ is relatively well constrained in the redshift range $z\in[0,5]$ by several observations \citep{Zwaan_2005, Rao_2006, Lah_2007, Songaila_2010, Martin_2010, Noterdaeme_2012, Braun_2012, Rhee_2013, Delhaize_2013, Crighton_2015}, the value of the HI bias is poorly known \citep{2007MNRAS.378..301B,2012ApJ...750...38M,Guo}. Thus, a better model of the HI bias would allow us to 1) improve our understanding of the astrophysical processes governing the abundance and evolution of HI across time, 2) improve the design of future 21cm experiments and optimize their main science cases and 3) produce more accurate forecasts for the constraining power of these observations. One of the indirect goals of this paper is to measure the HI bias at $z\approx 0$.

    In the absence of better data, the halo model \citep{Smith2003} offers an alternative method to predict the abundance and clustering of HI after including two extra ingredients: a relation between total halo mass and HI mass $\mhimh$, and a model for the distribution of HI within each halo $\rho_{\rm HI}(r|M_h)$. However, these extra degrees of freedom must be constrained using available data before this method can be useful to predict the cosmic HI signal. This has been done in the past by combining low-redshift data from HI surveys and column-density information from observations of the Lyman-$\alpha$ forest at higher redshifts \citep{Hamsa,2017MNRAS.471.1788C}, often revealing apparent tensions between datasets. In this paper we will use a self-consistent framework to constrain the $\mhi$-$M_h$ relation using the mass-weighed clustering of HI galaxies detected by the Arecibo Legacy Fast ALFA survey (ALFALFA), as well as their abundance in halos extracted from galaxy groups found in the SDSS galaxy survey. We will also explore the possibility of constraining the shape of the HI profile and the impact of modeling assumptions on our results.

    This paper is organized as follows. In section \ref{sec:theory} we describe the theoretical framework we use to characterize the abundance and clustering of HI. We outline the data employed in this work in section \ref{sec:data}. The methods used to analyze the data and compare with the theory predictions are illustrated in section \ref{sec:method}. The main results of this work are shown in section \ref{sec:results}. We discuss the results and summarize the conclusions of this work in section \ref{sec:discussion}.

  \section{HI halo model}\label{sec:theory}
    Numerical simulations show that almost all the HI in the post-reionization Universe is inside dark matter halos \citep{Paco_18,Paco_14}. Thus, one can use the halo-model formalism \citep{Smith2003,CooraySheth} to study the abundance and clustering of cosmic neutral hydrogen. The purpose of this paper is to constrain the HI-mass-to-halo-mass relation $\mhimh$ from direct measurements in selected galaxy groups, as well as from the clustering of HI sources. Extending the halo model to predict the properties of HI requires additional assumptions about the relation between the HI mass and the halo mass as well as the distribution of HI itself inside halos. We follow a prescription similar to that developed recently by \citep{Hamsa,2017MNRAS.471.1788C}. 

    We start by assuming that, on average, the HI content of halos depends solely on their mass, and we parametrize the $\mhimh$ relation as \citep{Paco_18,2017MNRAS.471.1788C,Hamsa}:
    \begin{equation} \label{eq:MHIM}
      \mhi(M_h)=M_0\,\left(\frac{M_h}{M_{\rm min}}\right)^\alpha\exp\left(-\frac{M_{\rm min}}{M_h}\right).
    \end{equation}
    In this model, the overall normalization $M_0$ can be immediately associated with the cosmic HI fraction $\Omega_{\rm HI}\equiv\rho_{\rm HI}/\rho_c$ at $z=0$, where $\rho_c$ is the critical density. Both quantities are related through:
    \begin{equation}
      \Omega_{\rm HI}\equiv\frac{\bar{\rho}_{\rm HI}}{\rho_{\rm c}}=\frac{1}{\rho_{\rm c}}\int_0^\infty dM_h\,n(M_h)\,\mhimh,
    \end{equation}
    where $n(M_h)$ is the halo mass function. The other two free parameters of the model are $\alpha$, which describes the scaling of $\mhi$ with halo mass, and the low-mass cutoff $M_{\rm min}$, which represents the threshold mass needed for a halo to host HI. This mass cut-off is expected, since the gravitational potential of small halos is not deep enough to trigger the clustering and cooling of the hot gas heated by the UV background \citep{Paco_18}.

    On small scales, the clustering of HI is dominated by its distribution within the halo (i.e. the so-called 1-halo term). Although our constraints will be based solely on the shape of the correlation function on larger scales, we use two different models for the HI density profile, in order to quantify the effect of this assumption on the final results:
    \begin{itemize}
      \item {\bf Altered NFW profile:} this is the model introduced and used in \cite{MallerBullock,BarnesHaehnelt2014,Hamsa,Paco_18} and assumes the radial profile of the form:
            \begin{equation}
              \rho_{\rm HI}(r|M_h)\propto(r+3/4r_s)^{-1}(r+r_s)^{-2}
            \end{equation}
            where $r_\mathrm{s}$ is the scale radius of the HI cloud, and is related to the halo virial radius $R_\mathrm{v}(M_h)$ by the concentration parameter -- $c_\mathrm{HI}(M_h,z)\equiv R_\mathrm{v}(M_h)/r_\mathrm{s}$. We follow \cite{Bullock,Maccio} and use a mass-dependent concentration parameter given by:
            \begin{equation}
              c_\mathrm{HI}(M_h,z=0) = 4\,c_{\mathrm{HI},0}\left(\frac{M_h}{10^{11} M_\odot}\right)^{-0.109}.
            \end{equation}.
      \item {\bf Exponential profile:} this is the model implemented in \cite{Hamsa}, and given by
            \begin{equation}
              \rho_{\rm HI}(r|M_h)\propto\exp{(-r/r_{\rm s})},
            \end{equation}       
    \end{itemize}
    In both cases the proportionality factors are automatically fixed by requiring that the HI mass be given by the volume integral of the density profile up to the halo virial radius $R_{\rm v}(M)$.
    \begin{equation}
      \mhi(M_h)=4\pi\int_0^{R_v}dr\,r^2\rho_{\rm HI}(r|M_h).
    \end{equation}
    Thus, both profiles are described by one additional free parameter, $c_{{\rm HI},0}$. The normalized HI density profile in Fourier space for the altered NFW profile is given in \cite{Hamsa} (see their Eq. A3), while the exponential profile is simply
    \begin{equation}
      u_{\rm HI}(k|M_h)= \frac{1}{(1+k^2r_{\rm s}^2)^2}.
    \end{equation}.

    The halo model prediction \citep{2017MNRAS.471.1788C,Hamsa,Paco_18} for the HI power spectrum, is given by the sum of a 1-halo and a 2-halo term:
    \begin{align}\label{eq:P12h}
      &P_{\rm HI,1h}(k) = F^0_2(k), \hspace{12pt} P_{\rm HI,2h}(k) = P_{\rm lin}(k) \left[F^1_1(k)\right]^2,\\\label{eq:hmff}
      &F^\alpha_\beta(k)\equiv\int n(M_h)b^\alpha(M_h)\left[\frac{\mhimh}{\bar{\rho}_{\rm HI}}\,u_{\rm HI}(k|M_h)\right]^\beta dM_h,
    \end{align}
    where $n(M_h)$ is the halo mass function, $b(M_h)$ is the halo bias and $P_{\rm lin}(k)$ is the linear matter power spectrum. For the halo mass function and bias, we use the parametrizations of \cite{Tinker2010}, derived from numerical simulations, and we adhere to halo masses defined by a spherical overdensity parameter $\Delta=180$
    \begin{equation}
      M_h=\frac{4\pi}{3}\rho_c\Omega_{\rm m}\Delta R_{\rm v}^3.
    \end{equation}

    Finally, our basic clustering data vector is the 2D projected correlation, given by the projection of  the 3D correlation function along the line of sight. This can be computed directly from the power spectrum as:
    \begin{align}\nonumber
      \Xi(\sigma)&=\int_{-\infty}^{\infty}d\pi\,\xi(\pi,\sigma)\\
      &=\int_0^\infty \frac{k\,dk}{2\pi}\left[P_{\rm HI,1h}(k)+P_{\rm HI,2h}(k)\right]\, J_0(k\sigma),
    \end{align}
    where $J_0(x)$ is the order-0 cylindrical Bessel function. To accelerate the computation of $\Xi(\sigma)$ we made use of FFTLog \cite{FFTLOG}.

    Our theoretical model therefore depends on four free parameters $\theta=\{M_0,M_{\rm min},\alpha,c_{\rm HI,0}\}$. We fix all cosmological parameters to values compatible with the latest $\Lambda$ Cold Dark Matter constraints measured by Planck \cite{2016A&A...594A..13P} ($H_0=70\,{\rm km\,s^{-1}Mpc^{-1}}$, $\Omega_{\rm m}=0.3075$, $n_s=0.9667$, $\sigma_8=0.8159$)\footnote{We fix the expansion rate to $h=0.7$ instead of its best-fit measurement $h=0.6774$ to match the choice in made in \cite{2018MNRAS.tmp..502J} to measure $\Omega_{\rm HI}$. We will report our final results as a function of $h_{\rm 70}\equiv H_0/70 \,{\rm km\,s^{-1}Mpc^{-1}}$}.

  \section{Data}\label{sec:data}
    \subsection{The $\alpha$.100 dataset}\label{ssec:data.alpha100}
      \begin{figure*}
        \centering
        \includegraphics[scale = 0.8]{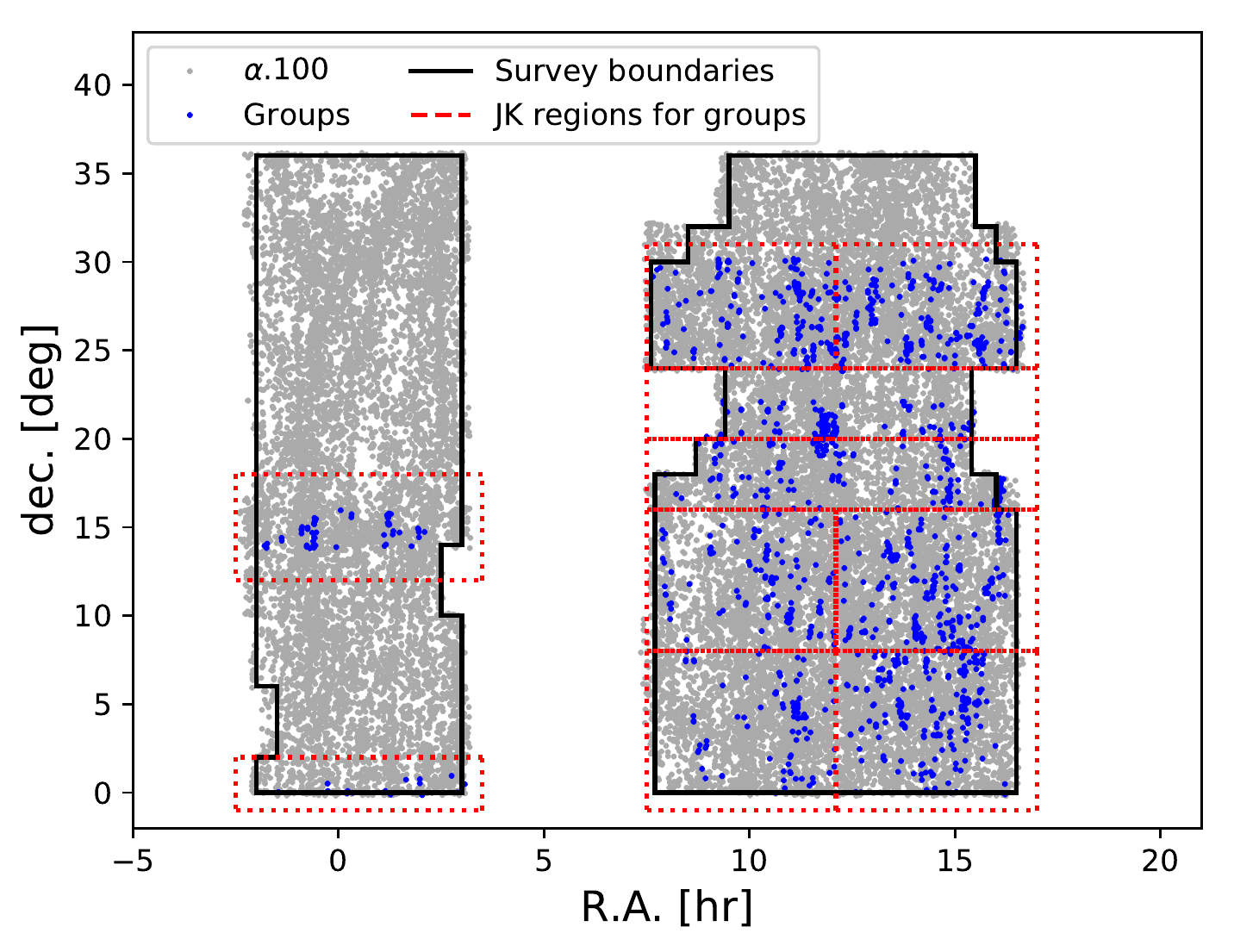}
        \caption{Sky distribution of the HI selected galaxies from $\alpha.100$ sample (gray dots). The black lines show the survey boundaries used in our clustering analysis (in which all sources outside the boundaries were omitted). The HI sources associated with groups in the SDSS DR7 group catalog are highlighted in blue. The dotted red lines show the jackknife regions used to estimate the cosmic variance uncertainties of the HI mass function in groups (see Section \ref{ssec:method.hicont}).}\label{fig:DATA}
      \end{figure*}
      The Arecibo Fast Legacy ALFA (Arecibo L-band Feed Array) survey, or ALFALFA{\footnote{\url{http://egg.astro.cornell.edu/alfalfa/}}} \citep{Giovanelli2005}, is a blind extragalactic HI survey performed using the Arecibo radio telescope. The main goal of ALFALFA is to quantify and study the properties of the HI content of the local Universe ($z\lesssim0.05$). It represents a significant improvement over previous HI surveys, with a beam FWHM of $\sim3.5$ arcmin, an rms noise of $\sim2.4\,{\rm mJy}$ and a spectral resolution of $\sim10\,{\rm km}\,{\rm s}^{-1}$. 
      
      Previous clustering analyses of the ALFALFA samples used the 40\% \citep{Haynes,Martin2012,Papastergis2013}, and 70\% \citep{Guo} data releases (labeled $\alpha$.40 and $\alpha$.70). Our analysis makes use of the final data release \citep{alfa100}, containing $\sim$31500 sources up to a redshift of $z = 0.06$ and covering approximately 7000 square degrees in two continuous regions at either side of the Galactic plane. Sources with good detection significance ($S/N>6.5$), classified as ``code-1'', represent the main sample ($\sim81\%$ of the total 31502 sources). Most of the remaining sources, classified as ``code-2'', correspond to lower signal-to-noise detections ($S/N>4.5$) with known optical counterparts. The remaining $\sim5\%$ of the catalog is mostly composed of high-velocity clouds of galactic provenance. We use only code-1 sources in the clustering analysis described in Section \ref{ssec:method.2pcf}, and both code-1 and code-2 objects in the direct measurement of the HI content of galaxy groups (Section \ref{ssec:method.hicont}). For each source, the catalog provides information about their angular coordinates, heliocentric radial velocity, radial velocity in the CMB frame, 21cm flux, line width and HI mass. HI masses for all objects can also be obtained from their distance and 21cm flux as
      \begin{equation}\label{eq:massobs}
        m_{\rm HI}=(2.356\times10^5 M_\odot)\,D^2\,S_{21}
      \end{equation}
      where $D$ is the distance to the source in Mpc, $S_{21}$ is the integrated flux in units of Jy km s$^{-1}$ and $m_{\rm HI}$ is the source's HI mass\footnote{To distinguish between the HI mass of ALFALFA sources and the total HI mass associated to a given dark matter halo, we label the latter $\mhi$ and the former $m_{\rm HI}$}.
      
      In the clustering analysis, the radial velocities $v_{\rm cmb}$ are used to assign radial distances to sources through their redshift $z_{\rm cmb}=v_{\rm cmb}/c$, using the cosmological parameters listed in Section \ref{sec:theory}. Due to the radio frequency interference (RFI) we make additional cuts and following \cite{Papastergis2013} we remove the sources outside $700\,{\rm km\,s^{-1}}<cz_{\rm cmb}<15000\,{\rm km\,s^{-1}}$. After performing these cuts in the raw data we are left with 24485 code-1 sources and 5365 code-2 sources. Figure \ref{fig:DATA} shows the angular distribution of all sources used in this work. The black lines delineate the survey boundaries used for in the clustering analysis. These cuts further reduce the clustering sample to 23438 objects. 
      
    \subsection{The SDSS group catalog}\label{ssec:data.groups}
      To assign the HI detected sources to dark matter halos, we cross-match the SDSS galaxies and the ALFALFA sources and determine the group membership of the cross-matched galaxies using a galaxy group catalog, following the procedure described in \cite{yoon_rosenberg_2015}. We use the SDSS DR7 group catalog\footnote{\url{http://gax.sjtu.edu.cn/data/Group.html}} updated from the DR4 group catalog \cite{yang_etal_2007}. The catalog uses galaxies in the SDSS DR7 spectroscopic sample with $0.01 \le z \le 0.2$ and redshift completeness $C>0.7$. The group finding algorithm has been extensively tested using mock galaxy redshift survey catalogs and has proven to be successful in associating galaxies that reside in a common halo \cite{yang_etal_2005}. In particular, this halo-based group finder works well for poor groups and identifies groups with only one member (i.e. isolated galaxies). The group halo masses are determined down to $M_h=10^{11.8} h^{-1} M_{\odot}$ using two methods: ranking by luminosity and from the stellar mass of member galaxies. Although we used the luminosity-ranked group halo mass, the results do not change if the stellar-mass-ranked halo mass is used instead. The group finder has been shown to correctly select more than 90\% of the true halos with $M_h \ge 10^{12} h^{-1} M_{\odot}$ \cite{yang_etal_2007}, which allows us to reliably study our galaxy samples within groups and clusters with halo mass $10^{12.50} h^{-1} M_{\odot} \le M_h \le 10^{15.04} h^{-1} M_{\odot}$.
      
      For the virial radius of groups with halo mass $M_h$, we adopt the radius $R_{180}$ that encloses an overdensity $\Delta=180$ times larger than mean density \citep{yang_etal_2007}:
      \begin{equation}
      \label{eq:rvir} 
      R_{180}=1.26 h^{-1} \mbox{Mpc} \left(\frac{M_h}{10^{14}h^{-1} M_{\odot}}\right)^{1/3} (1+z_{\rm group})^{-1},
      \end{equation}
      which is based on the WMAP3 cosmological model parameters \cite{spergel_etal_2007}, $\Omega_m=0.238$, $\Omega_{\Lambda}=0.762$ and $H_0=100 h$ $\mbox{km s}^{-1}\mbox{Mpc}^{-1}$, where $h=0.73$. While these parameters differ slightly from those used in this study, this does not significantly impact the results at the low redshifts of our sample ($z<0.055$). We also note that the DR7 group catalog has significant overlap only with the 70\% ALFALFA data release, and therefore no new information is gained by using the complete ALFALFA sample ($\alpha$.100 dataset).
      
      Figure \ref{fig:DATA} shows, in blue, the ALFALFA sources identified as members in the group catalog, as well as the jackknife regions used to compute the cosmic variance uncertainties for our estimate of the HI mass function in groups (dotted red lines, see Section \ref{ssec:method.hicont}).

  \section{Method}\label{sec:method}
    We derive constraints on the HI content of dark matter halos by using the clustering properties of HI galaxies weighted by their HI content, as well as direct measurements of the HI content of galaxy groups. We describe the procedures used to compile these two data vectors and their associated covariances here.

    As discussed in Section \ref{sec:intro}, our main interest is to quantify the properties of the total HI density inhomogeneities, since these are the relevant proxy of the density fluctuations measured by 21cm intensity mapping. To do so, our main assumption will be that the properties of the full HI density field can be inferred from the properties of HI-selected sources as measured by ALFALFA when weighed by their HI mass. This simplifying assumption should be a good approximation as long as the sources detected by ALFALFA account for a significant portion of the total HI mass. The validity of this assumption can be quantified to some extent by examining the measurements of the HI mass function measured by the ALFALFA collaboration in \cite{2018MNRAS.tmp..502J}, extrapolating it below the detection limit. This calculation shows that, for a conservative threshold of $m_{\rm HI,lo}=10^8\,M_\odot$, less than 5\% of the total HI would lie in sources not observed by ALFALFA. Thus, assuming that the tilt of the HI mass function does not vary sharply on smaller masses, the contribution from diffuse or undetected sources to the observables considered here is negligible given the uncertainties in our measurements. This is even more so for measurements of the HI clustering, given that the clustering bias of HI sources has been show to be only weakly dependent on HI mass \cite{Papastergis2013}. Even in the case of the measurement of the HI content in galaxy groups (see Section \ref{ssec:method.hicont}), where this contribution can rise to $\lesssim30\%$ we will explicitly show that the impact of the missing HI mass on our results is minimal.

    \subsection{The projected 2-point correlation function}\label{ssec:method.2pcf}
      Previous studies \citep{Martin2012,Papastergis2013,Guo} have measured two-point correlation function (2PCF) of HI-selected galaxies to determine their relation with the underlying dark matter density field. These studies have found that this sample has a low value of the clustering amplitude compared to the dark matter field (i.e.\ HI-selected galaxies have a low bias - $b_{\rm HI,g}$). Under the assumption described above, the same measurement can be performed on the 2PCF of HI-selected galaxies weighed by their HI mass to obtain a measurement of the total HI bias $b_{\rm HI}$, which plays a key role on 21cm intensity mapping studies. We describe the procedure used to estimate the 2PCF and its uncertainty here.

      We begin by estimating the 2D 2PCF $\xi(\pi,\sigma)$ as a function of the distance between pairs of objects along the line of sight ($\pi$) and in the transverse direction ($\sigma$). For this we use the Landy \& Szalay estimator \cite{LS}, given by
      \begin{equation}
        \xi(\pi,\sigma)=\frac{{\rm DD}(\pi,\sigma)-2{\rm DR}(\pi,\sigma)+{\rm RR}(\pi,\sigma)}{{\rm RR}(\pi,\sigma)},
      \end{equation}
      where ${\rm DD}$ is the normalized histogram of unique  weighted pairs of sources separated by a distance $(\pi,\sigma)$ found in the data catalog:
      \begin{equation}\nonumber
        {\rm DD}(\pi,\sigma)=\frac{\sum_{i=1}^N\sum_{j>i}w_iw_j\,\Theta(\pi_{ij};\pi,\Delta\pi)\,\Theta(\sigma_{ij}';\sigma,\Delta\sigma)}{\sum_{i=1}^N\sum_{j>i}w_iw_j}.
      \end{equation}
      Here $\pi_{ij}$ is the distance between the $i$-th and $j$-th objects along the line of sight (and similarly for the transverse distance $\sigma_{ij}$), and $\Theta(x\in(x_1,x_2))=1$ when $x\in(x_1,x_2)$ and $0$ otherwise. RR is defined similarly  for unique pairs of objects belonging to a random catalog with statistical properties similar to those of the data (e.g. in terms of spatial and weights distribution) but no intrinsic clustering. Finally, DR is given by all pairs of data-random objects. The weights $w_i$ assigned to each object are described below.

      \subsubsection*{Random catalog}
        The random catalog needed to compute the correlation function should follow the same redshift, angular and weights distribution observed in the data. We use the area cuts reported in \cite{2018MNRAS.tmp..502J} to define the survey footprint These are shown in Figure \ref{fig:DATA}, as black lines, and we discard all sources outside these boundaries. The angular positions of the random objects are then generated by drawing random coordinates with a constant surface density within this area.

\begin{figure*}
          \centering
          \includegraphics[width=0.49\textwidth]{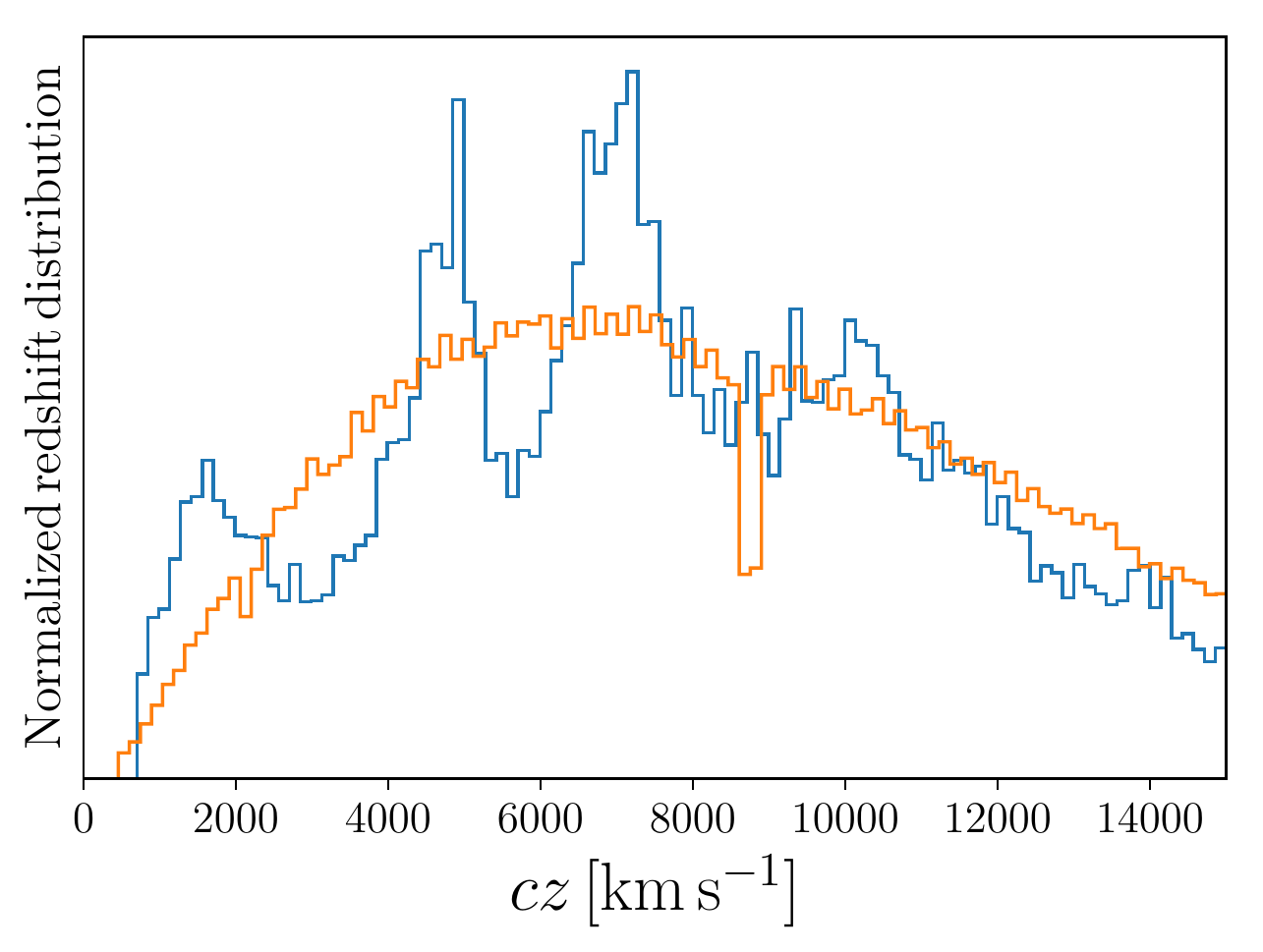}
          \includegraphics[width=0.49\textwidth]{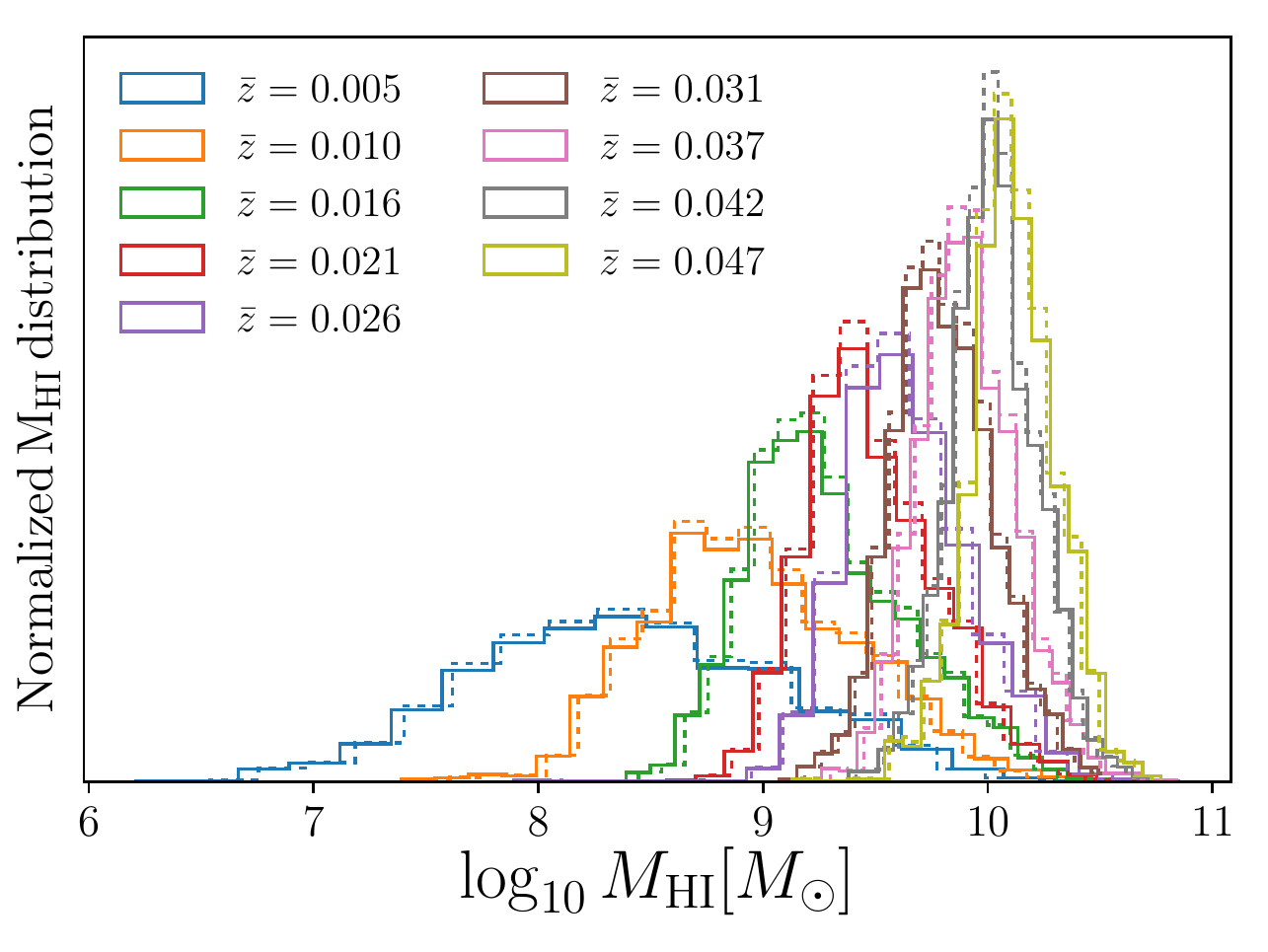}
          \caption{{\sl Left:} normalized redshift distribution in the data (blue) and the constructed random catalog (orange). {\sl Right:} the HI mass distribution in the data (solid line) and the constructed random catalog (dashed line) in different redshift bins (see legend).}\label{fig:Redshift_distribution}
        \end{figure*}

        We assign redshifts to the random objects by accounting for both the radial selection function described in \citep{Papastergis2013} (see their Figure 4) and for RFI incompleteness, using the completeness function presented in the same paper (see their Figure 6). Including these two effects is achieved by keeping a point with distance $d$ in the random catalog with a probability corresponding to the product of the selection and RFI completeness functions at $d$. The final normalized redshift distribution in both the data and the random catalog is shown in the left panel of Figure \ref{fig:Redshift_distribution}. 

        The points in the random catalog must also be assigned mass weights following the same $m_{\rm HI}$ distribution as the data. To achieve this, we split the random and the data set in 10 redshift bins. In each redshift bin we give each random point an HI mass randomly sampled from the data in the same bin. The resulting HI mass distributions are shown in the right panel of Figure \ref{fig:Redshift_distribution}.

      \subsubsection*{Weights}
\begin{figure*}
          \centering
          \includegraphics[width=0.9\textwidth]{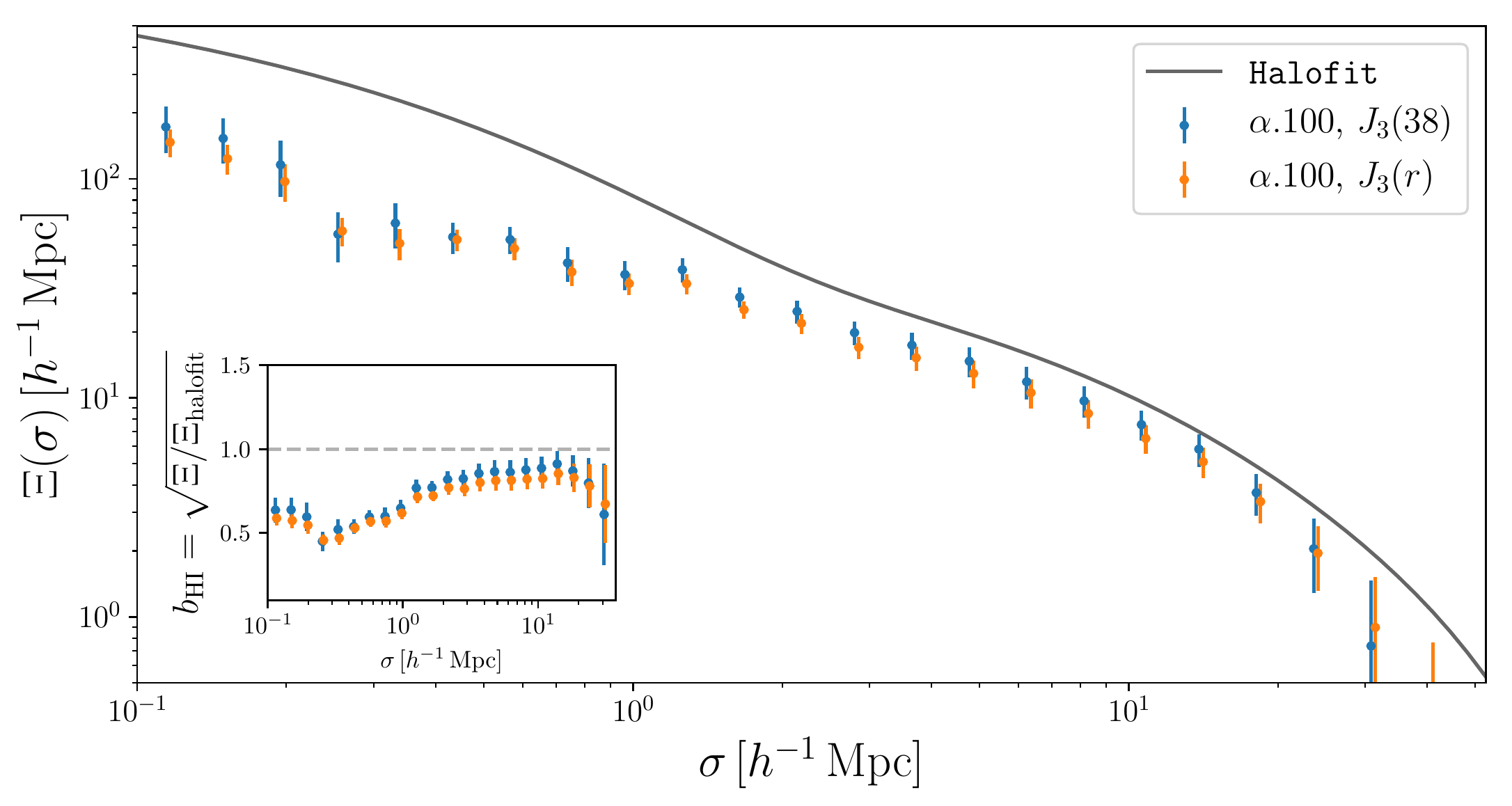}
          \caption{2D projected correlation function. The points with error bars show $m_{\rm HI}$-weighted correlation function computed for the $\alpha.100$ data set, while the black solid line shows the \texttt{HaloFit} prediction for the matter correlation function at $z=0$. Orange points show the measurements using pair-wise weights that depend explicitly on the pair separation (Eq. \ref{eq:weights}), while the blue points correspond to the case of fixing $J_3(r)$ to $J_3(38\Mpc)$, independent of separation. The impact of the choice of weighting scheme is found to be negligible. The inset shows the scale-dependent HI bias $b_{\rm HI}$ as a ratio of the measurement with respect to the matter correlation function. Orange points have been slightly shifted to the right.}\label{fig:Xi_sigma_jack}
        \end{figure*}

        The sample we use is not volume-limited, and the objects near the peak of the selection function will dominate the measured correlation function. In order to avoid this, we apply optimal pair-wise weights $w_{i,j}=w_i\times w_j$, where $w_i$ is given by \cite{1980lssu.book.....P,1994ApJ...426...23F}
        \begin{equation}\label{eq:weights}
          w_i=\frac{m_{\rm HI,i}}{1+4\pi n(d_i) J_3(r_{ij})},
        \end{equation}
        where $n(d_i)$ is the number density of the sample at the distance $d_i$ to the $i$-th source, $r_{ij}$ is the comoving separation between both objects and $J_3$ is an integral over the real-space isotropic correlation function:
        \begin{equation}
          J_3(r)=\int_{0}^{r}r'^2\,\xi(r')dr'.
        \end{equation}
        Implementing these weights requires an assumption about the shape and amplitude of $\xi(r)$. For these we follow \cite{Martin2012} and use $\xi(r)=(r/r_*)^{-1.51}$, with $r_*=3.3\Mpc$. In fact, we find that fixing $J_3(r)$ to $J_3(r=38\Mpc)=2962\,{\rm Mpc^3}$ is enough to obtain a close-to-optimal correlation function (see Fig. \ref{fig:Xi_sigma_jack}). When implementing these weights we approximated the number density as $n(d)=n_0\exp{(-(d/d_0)^\gamma)}$ where $n_0=0.23(\Mpc)^{-3}$, $d_0=31.18\Mpc$ and $\gamma=0.99$. These numbers were obtained by fitting the distance distribution of objects in the random catalog. Note also that Eq. \ref{eq:weights} already includes the $m_{\rm HI}$ weights needed to recover the clustering properties of the total HI density.

        Using this formalism, the measurement of the correlation function was carried out using the code \texttt{CUTE} \citep{CUTE}. We adopted a logarithmic binning in $\sigma$ in the range $\sigma\in[0.11,52)\Mpc$ with $\Delta\log_{10}\sigma/(h^{-1}\,{\rm Mpc})=0.12$, and we used 59 linear bins of $\pi$ in the range $\pi\in[0.5,59.5)$ $\Mpc$. In order to eliminate the effect of redshift-space distortions and be able to compare our measurements with the real-space theoretical prediction, we compute the projected correlation function $\Xi(\sigma)$ by integrating $\xi(\pi,\sigma)$ along the line of sight:
        \begin{equation}
          \Xi(\sigma)=\int_{-\infty}^\infty d\pi\,\xi(\sigma,\pi)\simeq2\sum_0^{\pi_{\rm max}}\xi(\sigma,\pi)\Delta\pi,
        \end{equation}
        where, as in \cite{Martin2012}, we used $\pi_{\rm max}=30\Mpc$.

        Figure \ref{fig:Xi_sigma_jack} shows the measured HI-mass-weighted, projected correlation function (points with error bars) together with the prediction for the projected correlation function of the total matter overdensity, obtained from the HaloFit model for the matter power spectrum \cite{2012ApJ...761..152T}. The scale-dependent HI bias is shown in the inset of the same figure as the square root of the ratio of both quantities. The measured $b_{\rm HI}$ is in good agreement with the measurement of the bias of HI-selected galaxies presented in \cite{Martin2012}. This is to be expected, given the observation that the clustering of HI sources shows little or no dependence on HI mass\footnote{Note that \cite{Guo} observe a significant dependence on HI mass above $10^9\,M_\odot$. This possible dependence at high masses, however, does not alter our assumption that the ALFALFA sources can be used to study the properties of the overall HI distribution, including all structures below ALFALFA's detection limit.}.

      \subsubsection*{Covariance matrix}
\begin{figure}
          \centering
          \includegraphics[width=0.49\textwidth]{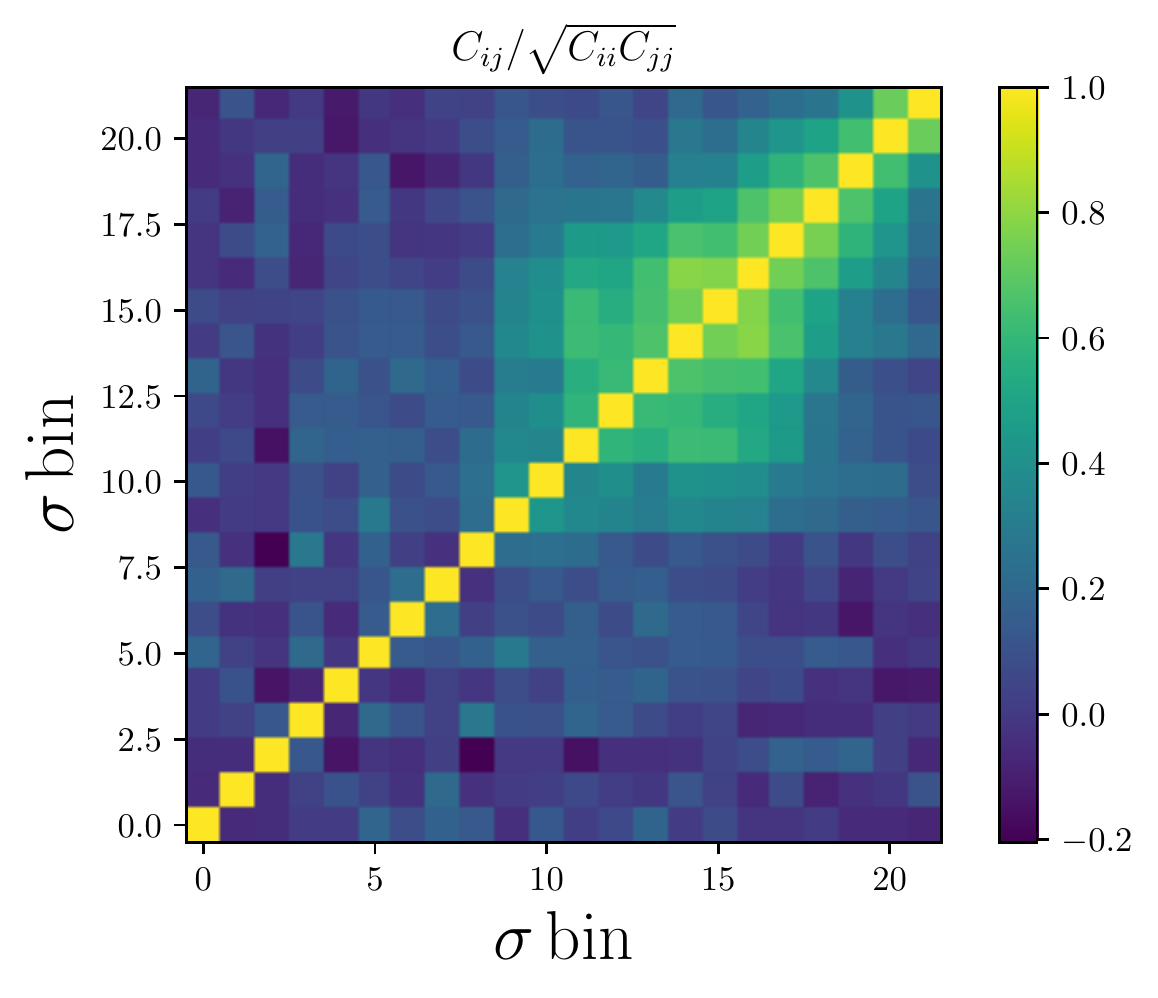}
          \caption{Jackknife correlation matrix for the projected 2-point correlation function. We use 22 logarithmic bins in the transverse separation $\sigma$ in the range $\sigma\in[0.11,30.8)\,h^{-1}{\rm Mpc}$.}\label{fig:Cov_mat}
        \end{figure}

        We estimate the uncertainties on the measured projected correlation function using the jackknife resampling method \citep{Jackknife,Zehavi2002}. We divide the survey footprint into $N=156$ contiguous patches covering $\sim40\deg^2$ each. We remove one patch at a time and measure the projected correlation function in the remaining area. The jackknife estimate of the covariance matrix is then given by:
        \begin{equation}
          C_{ij}={\rm Cov}(\Xi_i,\Xi_j)=\frac{N_s-1}{N_s}\sum_{p=0}^{N_s}(\Xi_i^p-\bar{\Xi}_i)(\Xi_j^p-\bar{\Xi}_j).
        \end{equation}
        Here $\Xi^p_i$ is the correlation function measured in the $i$-th bin after omitting the $p$-th patch and $\bar{\Xi}_i$ is the average of $\Xi^p_i$ over all patches. Figure \ref{fig:Cov_mat} shows resulting correlation matrix $r_{ij}=C_{ij}/\sqrt{C_{ii}C_{jj}}$.

        Ultimately we are interested in the inverse covariance matrix. The inverse of the jackknife covariance is a biased estimate of the true inverse covariance, and we correct for this bias with an overall normalization factor \citep{Hartlap_2007}:
        \begin{equation}
          C^{-1}\rightarrow\frac{N_{\rm s}-N_{\rm b}-2}{N_{\rm s}-1}\,C^{-1},
        \end{equation}
        where $N_{\rm s}=156$ is the number of jackknife samples and $N_{\rm b}=22$ is the number of $\sigma$ bins used in the analysis.

    \subsection{HI content in groups}\label{ssec:method.hicont}
\begin{figure}
        \centering
        \includegraphics[width=0.47\textwidth]{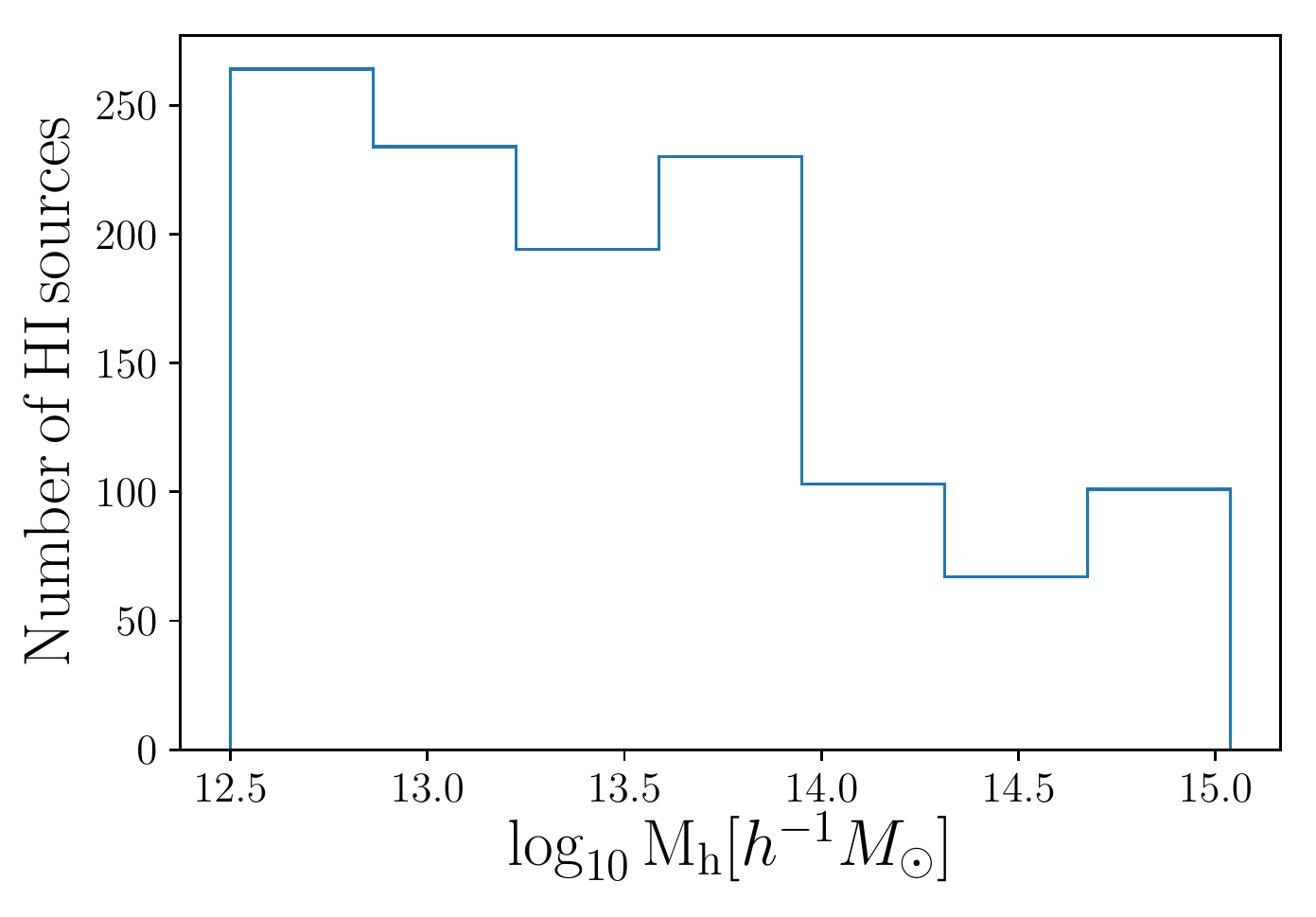}
        \includegraphics[width=0.47\textwidth]{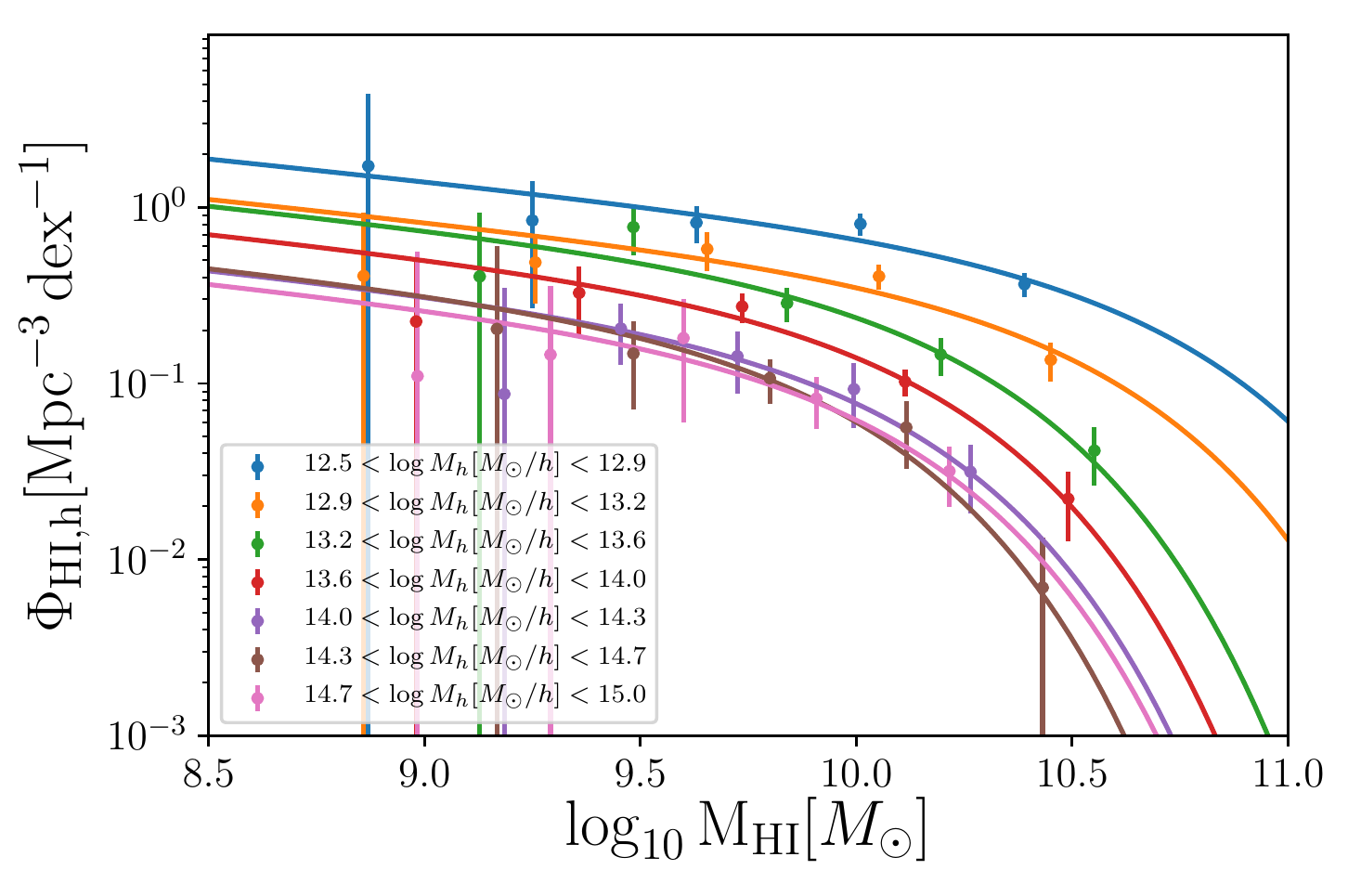}
        \caption{{\sl Top:} the number of HI sources in the SDSS group catalog lying in each halo mass bin after RFI and 50\% completeness cuts. {\sl Bottom:} HI mass functions estimated from the SDSS group catalog using 2DSWML method in different halo mass bins (see legend).}
        \label{fig:HIMF_halos}
      \end{figure}

      As described in Section \ref{ssec:data.groups}, we also include direct constraints on the $\mhimh$ relation in our analysis, coming from the matching of ALFALFA sources to optical members of galaxy groups with calibrated halo mass detected in the SDSS group catalog. To minimize a potential bias due to the incomplete coverage of the sky-projected area for each group, we estimate volume-correction factors for few large groups near the ALFALFA survey boundary. An estimate of the HI mass of each group is made by directly summing the masses of all ALFALFA member sources and applying the corresponding area correction factor, which is almost negligible for most of the groups. In general, this estimate of the group HI mass would be biased low, since the estimator will miss all ALFALFA sources with no optical counterparts lying in the comoving volume of each group, as well as any diffuse or unresolved HI component. The first cause of this bias (the sources with no optical detections) should have a negligible impact on this study, since it affects only $\sim6\%$ \citep{2011AJ....142..170H} of all ALFALFA sources, and most of those are expected to be galactic high-velocity clouds, and not extragalactic in nature. To quantify and minimize the impact of contributions from undetected HI components, we estimate the HI mass function (i.e. the $m_{\rm HI}$ distribution of ALFALFA sources) in bins of group halo mass. The exact procedure is as follows:
      \begin{enumerate}
        \item We separate the SDSS group catalog into $7$ logarithmically spaced bins of halo mass in the interval $\log_{10}M_h/(h^{-1}M_\odot) \in[12.50,15.04]$. The top panel of Figure \ref{fig:HIMF_halos} shows the number of HI sources lying in each of these mass bins.
        \item In each bin we estimate the HI mass function $\phi(m_{\rm HI})$ using all the member sources found in the ALFALFA dataset. For this we use the 2D step-wise maximum likelihood (2DSWML) estimator described below.
        \item In order to extrapolate below the detection limit, we model the measured mass function as a Schechter function with the form
        \begin{equation}
          \phi(m_{\rm HI})=\ln(10)\,\phi_*\left(\frac{m_{\rm HI}}{M_*}\right)^{\alpha_s+1}\exp\left(-\frac{m_{\rm HI}}{M_*}\right).
        \end{equation}
        \item For each halo mass bin, we compute the corresponding HI mass (and its uncertainty) by integrating over the reconstructed HI mass function, propagating all uncertainties as described below. We also compute a second estimate of the HI mass by integrating over the measured, model-independent 2DWSML mass function. This can only be done within the range of HI masses covered by ALFALFA, and the comparison of these two estimates then allows us to quantify the systematic uncertainty associated with undetected HI sources.
      \end{enumerate}
      The list of reconstructed HI masses as a function of group halo mass is then appended to the correlation function described in the previous section to form the total data vector.      

      \subsubsection*{The 2DSWML mass function estimator}
        The idea of step-wise maximum-likelihood estimators has been applied in the past to reconstruct the luminosity function from a magnitude-limited sample \citep{1988MNRAS.232..431E,2001MNRAS.326..255C,2016MNRAS.457.4393J}. The method is non-parametric, modeling the luminosity function as sum of top-hat functions, and finding their amplitudes by maximizing the likelihood of the observed sample. The latter is possible interpreting the luminosity function as  a probability distribution. The same logic was applied by \cite{2010ApJ...723.1359M,2018MNRAS.tmp..502J} to estimate the HI mass function of ALFALFA sources, with the added complication that the completeness of the sample depends on both the HI flux $S_{21}$ and the 21cm line width $W_{50}$. This gives rise to the 2-dimensional step-wise maximum-likelihood estimator (2DSWML), which we describe briefly here. To simplify the notation we will define here $\mu\equiv\log_{10}m_{\rm HI}/M_\odot$ and $w\equiv\log_{10}W_{50}/{\rm km}\,{\rm s}^{-1}$.
        
        The probability that a source $g$ is detected with mass $\mu_g$ and line width $w_g$ at distance $d_g$ (within an interval $\Delta\mu,\,\Delta w$) is given by
        \begin{equation}
          p_g=\frac{\phi(\mu_g,w_g)\,\Delta\mu\,\Delta w}{\int_{-\infty}^\infty dw\int_{\mu_{\rm lim}(d_g,w)}^\infty d\mu\,\phi(\mu,w)},
        \end{equation}
        where $\phi(\mu,w)$ is the joint distribution of HI masses and line widths. Let us now model $\phi(\mu,w)$ as a 2D step-wise function, taking constant values in intervals of $\mu$ and $w$. Then, maximizing the log-likelihood $\mathcal{L}=\prod_gp_g$, we obtain an expression for the best-fit amplitudes $\phi_{i,j}$ in the $i$-th interval of $\mu$ and the $j$-th interval of $w$:
        \begin{equation}\label{eq:2dswml}
          \phi_{i,j}=n_{i,j}\,\left[\sum_g\frac{H_{g,ij}}{\sum_{i',j'}H_{g,i'j'}\phi_{i',j'}}\right]^{-1},
        \end{equation}
        where $g$ runs over all sources in the sample, $n_{i,j}$ is the number of galaxies in bin $(i,j)$ and $H_{g,ij}$ is the mean completeness of the sample in that bin for sources at a distance $d=d_g$. The completeness function was determined as described in \cite{2010ApJ...723.1359M}. We imposed a hard cut on $m$ and $w$, using only bins with completeness $>50\%$. We verified that our results did not vary significantly with more stringent completeness cuts.
        
        Note that Eq. \ref{eq:2dswml} gives $\phi_{i,j}$ recursively as a function of itself, and in practice $\phi_{i,j}$ is found through an iterative process. Once a converged solution for $\phi_{i,j}$ has been found, the HI mass function is obtained by marginalizing over $W_{50}$:
        \begin{equation}
          \phi_i=\sum_j \phi_{i,j}\Delta w.
        \end{equation}
        Finally, this method is able to determine $\phi_{i,j}$ up to an overall normalization constant. We fix this by matching the integral of $\phi(m_{\rm HI},W_{50})$ to the total number of ALFALFA sources in each halo mass bin divided by the comoving volume covered by the corresponding halos, as described in Appendix B of \cite{2010ApJ...723.1359M}.
        
        The bottom panel of Figure \ref{fig:HIMF_halos} shows the estimated HI mass functions in each halo mass bin used in this analysis, together with their best-fit Schechter models. For this figure, the mass functions were normalized dividing by the total volume enclosed within the virial radii of all groups in each halo mass bins. Note that, since we only use $\phi(m_{\rm HI})$ to estimate the $\mhimh$ relation, our results are independent of this volume, and only depend on the total number of HI sources and galaxy groups in each $M_h$ bin.

      \subsubsection*{Error propagation}
        \begin{figure*}
          \centering
          \includegraphics[width=0.7\textwidth]{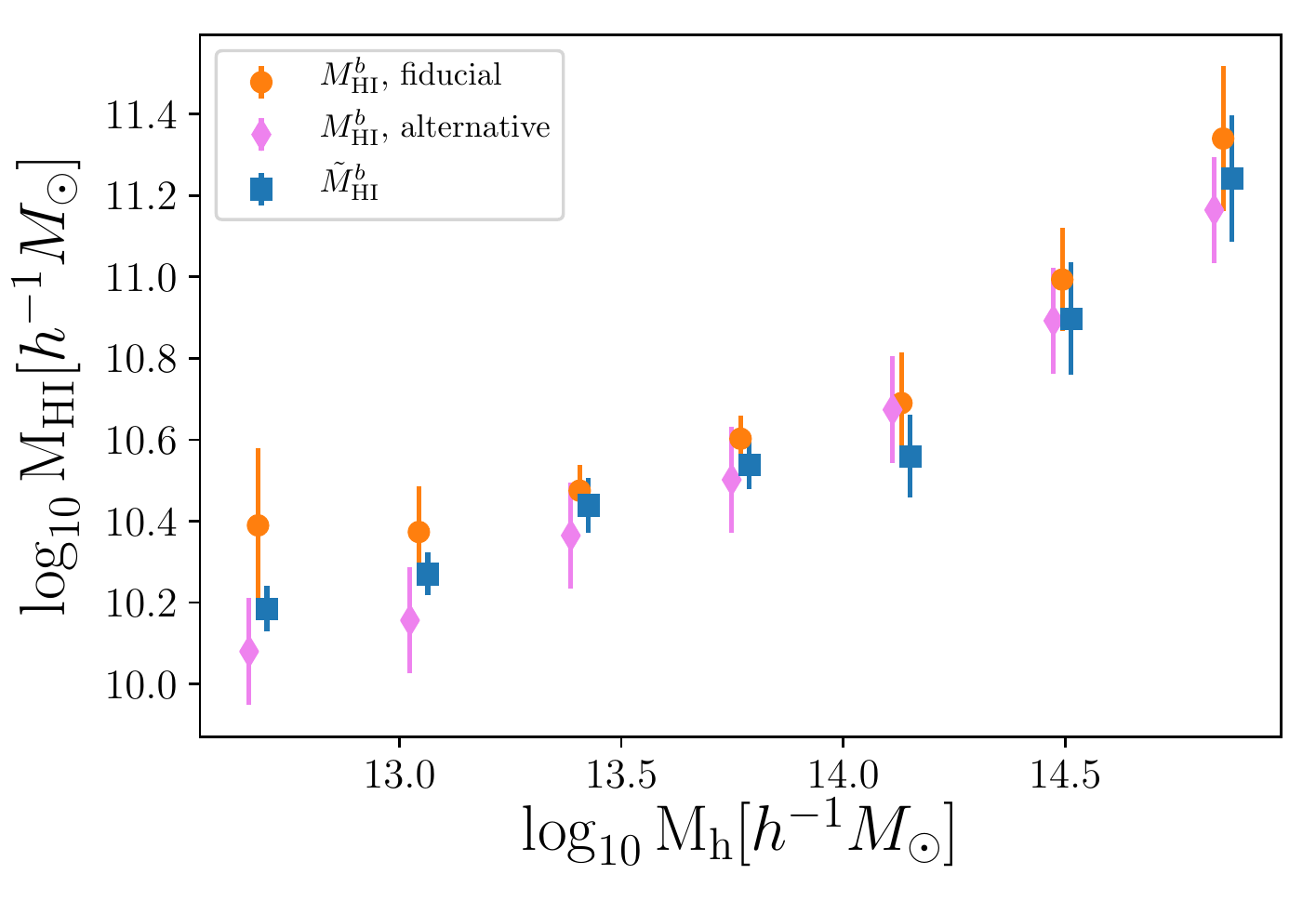}
          \caption{Estimated total $M_{\rm HI}$ in each halo mass bin obtained from the HI mass functions using three different methods. The total $M_{\rm HI}$ estimated by fitting the HI mass functions using the Schechter parametrization and accounting for the missing HI mass are shown with orange points. The error bars are computed by propagating the Schechter parameter uncertainties. The blue squares show the results of directly integrating the HI mass functions over the available range of HI masses, i.e.\ without extrapolation. The error bars in this case are computed by propagating the 2DSWML mass function uncertainties in quadrature. The violet diamonds show the second alternative estimate, found by rescaling the best-fit HI mass function in each halo mass bin (see Section \ref{ssec:method.hicont}). The corresponding error bars are computed from the uncertainties in the mass function found by \protect\cite{2018MNRAS.tmp..502J}.}
          \label{fig:MHIMF_comp}
        \end{figure*}
        The uncertainties in the $\mhimh$ relation inferred from the HI richness of groups, as described above, are driven by the errors in our estimate of the mass function in each $M_h$ bin.
        Four main sources of uncertainty contribute to these errors \citep{2018MNRAS.tmp..502J}, and we account for them as follows:
        \begin{enumerate}
          \item {\bf Poisson:} with each measurement of $\phi_i$ we associate a Poisson-counting error given by $\sigma(\phi_i)=\phi_i/\sqrt{N_i}$, where $N_i$ is the number of sources contributing to the $i$-th $m_{\rm HI}$ bin.
          \item {\bf Sample variance:} the uncertainty associated with the stochastic variation in $\phi_i$ induced by the particular density fluctuations covered by the survey volume of ALFALFA was quantified through the jackknife resampling method described in Section \ref{ssec:method.2pcf}. In this case we use the 10 jackknife regions shown as red dotted lines in Figure \ref{fig:DATA}.
          \item {\bf Mass measurement errors:} the HI mass of each source is inferred from its 21cm flux and its radial comoving distance. Both quantities have associated measurement uncertainties which propagate into $m_{\rm HI}$, shifting sources between different HI mass bins. To account for this, we generated 100 random realizations of the $\alpha.70$ catalog by adding a random Gaussian error to the distances and fluxes of all sources (with a standard deviation given by their estimated error). We re-computed the HI masses and corresponding $\phi(m_{\rm HI})$ for each realization (see Eq. \ref{eq:massobs}), and estimate the uncertainty associated to these errors from the scatter of all realizations.
          \item {\bf Line width measurement errors:} errors in $W_{50}$ also affect our measurement of the 2DSWML mass function, by shifting sources between different $W_{50}$ bins. The associated uncertainties were estimated from 100 random realizations, following the same procedure described above for mass measurement errors.
        \end{enumerate}
        We added the errors associated with these 4 sources in quadrature to find the final uncertainties on $\phi_i$.
        
        Once $\phi_i$ and its uncertainties have been measured, we find the best-fit Schechter models in each $M_h$ bin. To avoid over-fitting, given the relatively small number of points in which we estimate the mass function for each bin, we fix the tilt of the Schechter function to its best-fit value for the overall HI mass function as reported by \cite{2018MNRAS.tmp..502J}, $\alpha_s=-1.25$. The best-fit Schechter functions in each $M_h$ bin are shown as solid lines in Figure \ref{fig:HIMF_halos}.
        
        To estimate the uncertainties in the Schechter parameters $(\phi_*,M_*)$, we sample their likelihood running a Markov chain Monte Carlo (MCMC). For any point $(\phi_*,M_*)$ in these chains, the corresponding HI mass for halos in the $b$-th $M_h$ bin can be estimated as:
        \begin{align}\nonumber
          \mhi^b&=\frac{V_b}{N^b_{\rm group}}\int_0^\infty\phi(m_{\rm HI})\,m_{\rm HI}\,d\log_{10}m_{\rm HI}\\\label{eq:schechint}
                &=\frac{V_b\phi_*\,M_*}{N^b_{\rm group}}\,\Gamma\left(2+\alpha\right),
        \end{align}
        where $V_b$ is the uncorrected volume spanned by all groups in the $b$-th $M_h$ bin and $N^b_{\rm group}$ is the corresponding number of groups. Our final estimate of the $\mhimh$ relation (and its uncertainty) from the galaxy group data is then given by the mean of $\mhi$ (and its scatter) across all points in the MCMC chain. Finally, we correct our results for self-absorption as described in \cite{2018MNRAS.tmp..502J}. The results are shown as orange points with error bars in Fig. \ref{fig:MHIMF_comp}.

        Since our measurement of $\mhi^b$ involves extrapolating the HI mass function to very small masses, below the ALFALFA detection limit at the group's redshift, it is worth quantifying the impact of this extrapolation on our results. We do so here by comparing the fiducial measurement of $\mhi^b$ described above, with two alternative estimates:
        \begin{enumerate}
         \item The first estimator is given by directly integrating the measured 2DSWML mass function over the available range of HI masses in ALFALFA. Labeling the 2DSWML in the $b$-th halo mass bin as $\phi^b_i$, this alternative estimate is given by
         \begin{equation}
           \tilde{M}_{\rm HI}^b=\frac{V^b}{N_{\rm group}^b}\sum_i \phi^b_i\,10^{\mu_i}\,\Delta \mu.
         \end{equation}
         The uncertainty on $\tilde{M}_{\rm HI}^b$ can be estimated trivially from the uncertainties on $\phi^b_i$. Since $\tilde{M}^b_{\rm HI}$ and $\hat{\phi}^b_i$ are linearly related, the uncertainties on $\phi^b_i$, quantified as described above, can be propagated into $\tilde{M}_{\rm HI}^b$ in quadrature.
         \item The second estimator is produced by rescaling the best-fit HI mass function found by \cite{2018MNRAS.tmp..502J} in each halo mass bin. The rescaling factor for each group in the bin is estimated as the ratio of the observed number of sources found in that group to the number expected given the 2DSWML estimate of \cite{2018MNRAS.tmp..502J} accounting for sample completeness at the distance to the group. $M^b_{\rm HI}$ is then estimated by applying Eq. \ref{eq:schechint} to the Schechter function found in \cite{2018MNRAS.tmp..502J} rescaled by the factor above.
         
         Unlike our fiducial estimator, this alternative method has no free parameters, and can therefore be used to explore the possible consequences of over-fitting the per-bin mass functions based on a small number of objects. The main drawback of this estimator is that, by constructions, it assumes that the $m_{\rm HI}$ distribution in groups is the same as in the field.
        \end{enumerate}
        These alternative measurements of the $\mhimh$ relation are shown as blue squares and pink diamonds with error bars in Fig. \ref{fig:MHIMF_comp}.
        
        As could be expected, the measurements corresponding to the first alternative estimator are consistently below our fiducial estimates generated from the integral of the Schechter functions, with the missing mass corresponding to the contribution of sources below the ALFALFA detection limit. However, the associated mass difference is mostly below $\sim$25\% of our fiducial mass measurements throughout the full mass range. Since this offset is always smaller than the 1$\sigma$ statistical uncertainties, we find the impact of extrapolating the mass function to lower masses to be minimal. Note also that the blue error bars are consistently smaller than the orange ones. This is also to be expected, since the errors on $\tilde{M}^b_{\rm HI}$ estimated as described above, do not account for the additional uncertainty associated with mass below the detection limit.
        
        The second estimator, based on extrapolating the overall HI mass function, agrees well with our fiducial measurements in general, although it is noticeably lower in the two lowest $M_h$ bins. This is caused by the larger value of $M_*$ preferred by our Schechter fits in the low-$M_h$ bins. This result is consistent with previous measurements of the HI mass function around the region of the Virgo cluster, which suggest that massive ($\sim10^{15}M_\odot$) halos have a smaller $M_*$ than the field. Although this could be caused by ram pressure or tidal stripping, a better understanding of this result will require a more detailed study of the HI content in low-mass halos in both data and simulations \cite{Paco_18}. In any case, both estimates of $M^b_{\rm HI}$ are compatible within present uncertainties, and therefore we conclude that our measurements of this quantity are robust with respect to the method used to estimate it.

  \section{Results} \label{sec:results}
    \subsection{Fiducial results} \label{ssec:results.fiducial}
      We produce constraints on the three parameters of the $\mhimh$ relation (Eq. \ref{eq:MHIM}), $\theta\equiv\{\log_{10}M_0,\log_{10}M_{\rm min},\alpha\}$, from a joint data vector composed of three parts:
      \begin{enumerate}
        \item Measurements of the projected correlation function $\Xi(\sigma)$ (see Section \ref{ssec:method.2pcf}) in $N_\Xi=17$ logarithmic bins of $\sigma$ between $0.43\,h^{-1}{\rm Mpc}$ and $30.8\,h^{-1}{\rm Mpc}$. We use the altered NFW HI density profile described in Section \ref{sec:intro} as our fiducial model for the small-scale correlation function. We study the impact of this choice, as well as the choice of scale cuts in Section \ref{ssec:results.smallscales}.
        \item Direct measurements of the $\mhimh$ relation (see Section \ref{ssec:method.hicont}) in the $N_M=7$ logarithmic bins of halo mass shown in Fig. \ref{fig:MHIMF_comp}. Our fiducial measurements consist of the $\mhi$ estimates derived from the integral of the best-fit Schechter HI mass functions in each $M_h$ bin. We show the impact of extrapolating the HI mass function below ALFALFA's detection limit on our results in Section \ref{ssec:results.lowmass}.
        \item One measurement of the cosmic HI abundance $\Omega_{\rm HI}=(3.9\pm0.1\,({\rm stat.)}\pm0.6\,({\rm syst}))\times10^{-4}$ from ALFALFA's $\alpha$.100 sample, as reported by \cite{2018MNRAS.tmp..502J}. In terms of the halo model, the cosmic abundance receives contributions from the HI content of halos with arbitrarily small masses. Since our direct measurements of the $\mhimh$ relation do not go below $\log_{10}M_h/(h^{-1}{\rm Mpc})\simeq12.5$, this additional data point allows us to break the degeneracy between the overall amplitude $M_0$ and the minimum halo mass $M_{\rm min}$ of the $\mhimh$ relation.
      \end{enumerate}
      
      \begin{figure*}
        \centering
         \includegraphics[width=0.7\textwidth]{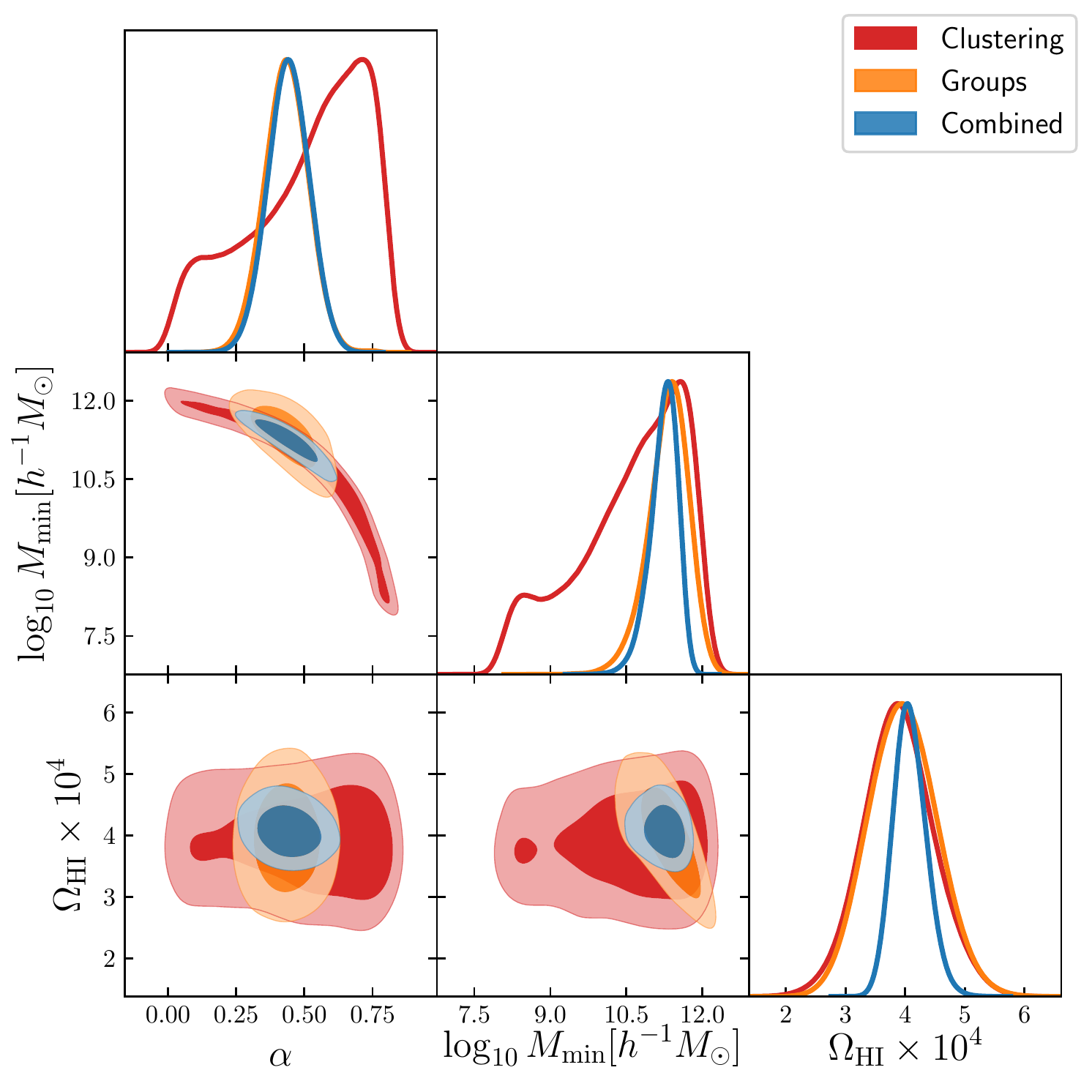}
        \caption{Final constraints on the parameters of the $\mhimh$ relation. Results are shown for the combination of clustering data and the $\Omega_{\rm HI}$ measurement (red), for the measurements of $\mhi$ in groups and $\Omega_{\rm HI}$ (light orange) and for the combination of the three datasets (blue).}
        \label{fig:triangle_fiducial}
      \end{figure*}
      \begin{figure*}
        \centering
        \includegraphics[width=0.9\textwidth]{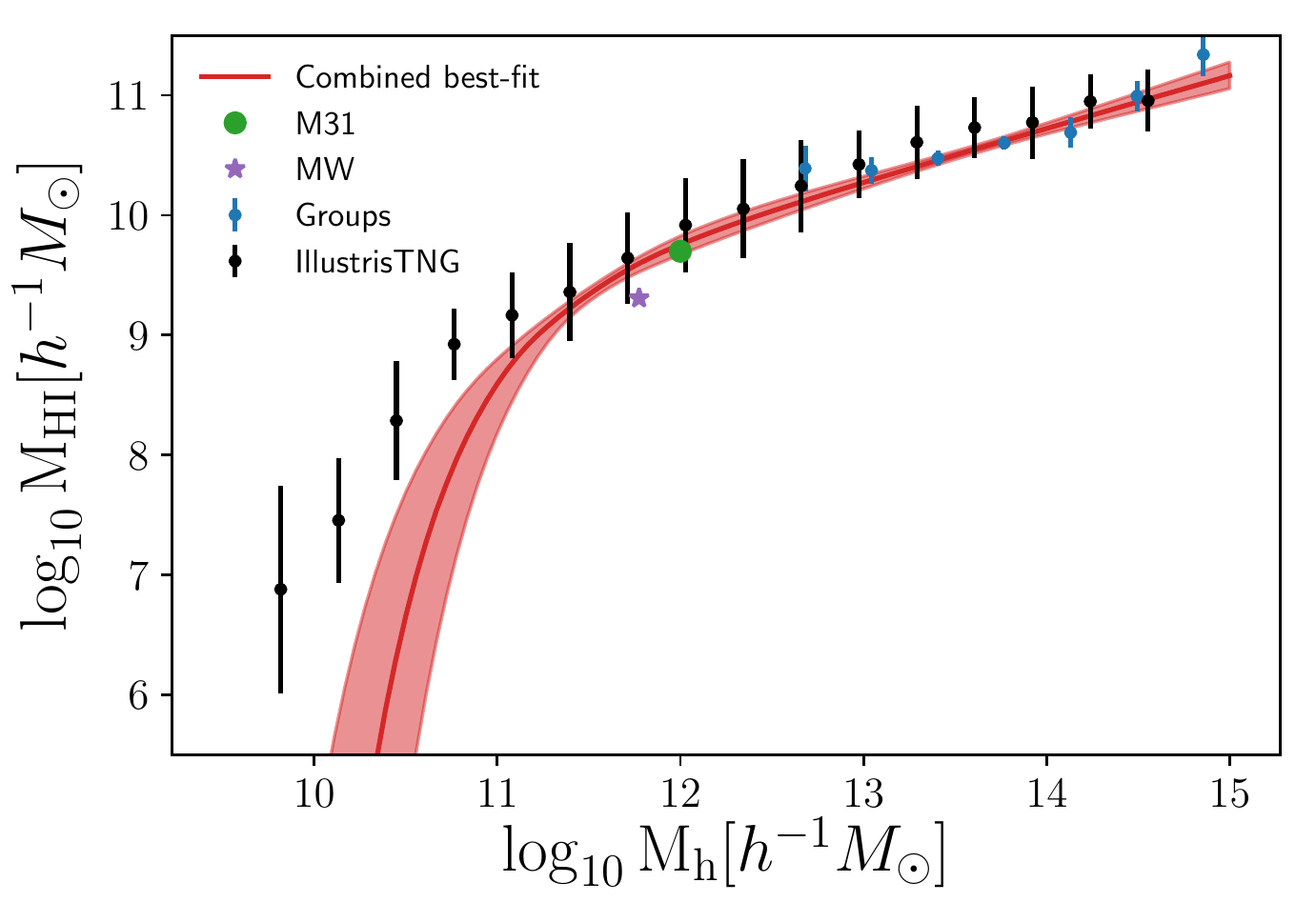}
        \caption{Combined best-fit $\mhimh$ relation (red solid line) together with the $1\sigma$ uncertainty (red shaded region) using three datasets: the projected mass-weighted correlation function $\Xi(\sigma)$, the direct estimates of the $\mhimh$ relation from the galaxy group catalog (shown also as blue points with error bars) and the measurement of the cosmic HI abundance $\Omega_{\rm HI}$ in \protect\cite{2018MNRAS.tmp..502J}. For comparison we show the results from the IllustrisTNG magneto-hydrodynamic simulation (black points) \citep{Paco_18} with the error bars corresponding to the typical per-halo scatter. We also show measurements of $M_h$ and $\mhi$ for individual galaxies: the Milky Way (purple star) and M31 (green circle).}\label{fig:mhimh_summary}
      \end{figure*}
      \begin{figure*}
        \centering
        \includegraphics[width=0.9\textwidth]{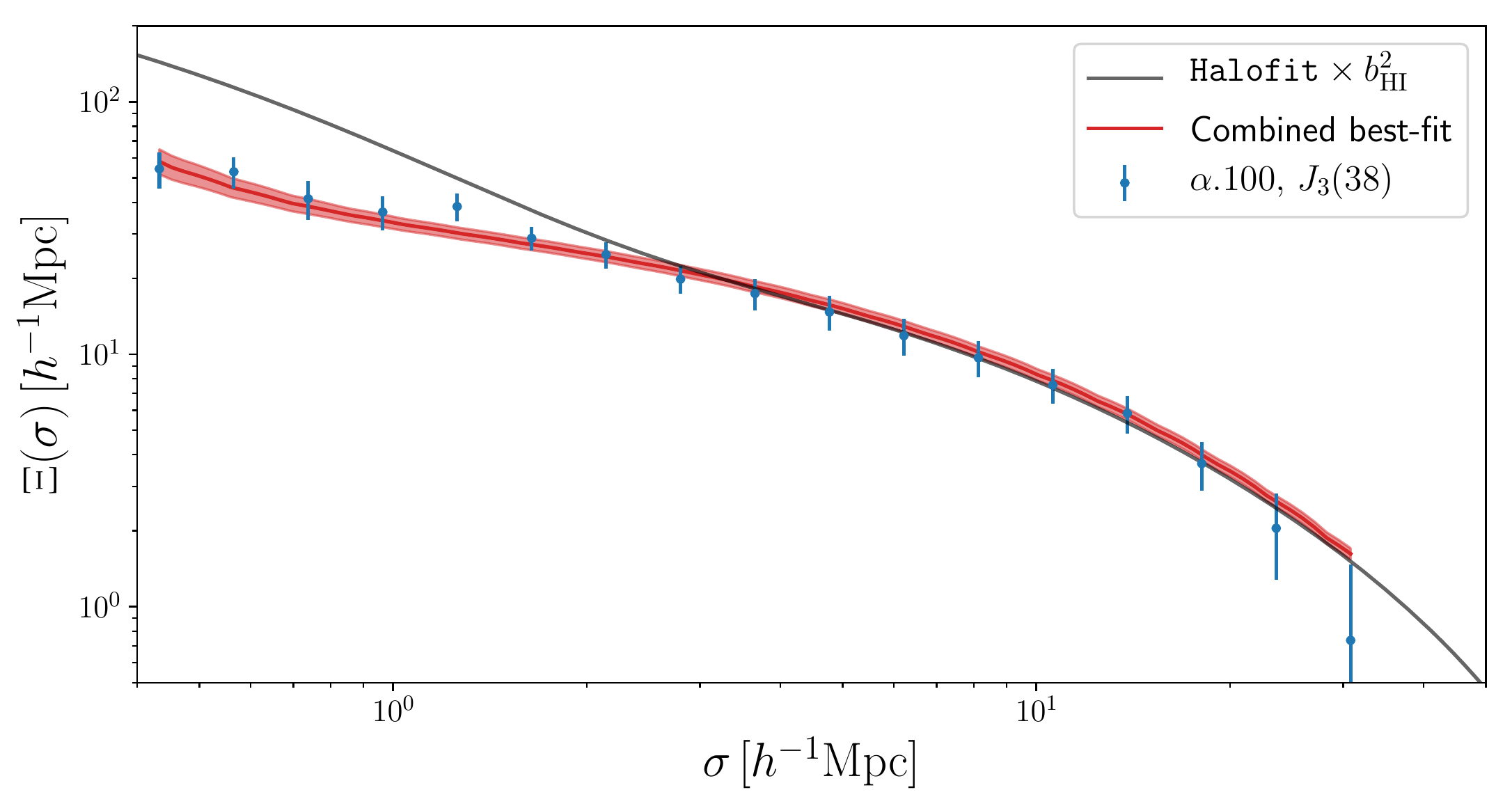}
        \caption{Predicted projected correlation function for our best-fit $\mhimh$ relation (red solid line) and its $1\sigma$ uncertainty (red shaded region). The blue points with error bars show the direct measurements from ALFALFA, and the black solid line corresponds to the HaloFit prediction for the matter correlation function scaled by our our best-fit $b_{\rm HI}^2$ (see Section \ref{sec:discussion}).}
        \label{fig:2pcf_best_fit}
      \end{figure*}
      Our fiducial data vector ${\bf d}$ therefore contains $N_\Xi+N_M+1=25$ elements, which we use to constrain the three-parameter model of the $\mhimh$ relation (in addition to the profile concentration parameter $c_{\rm HI,0}$, which we marginalize over). Assuming Gaussian statistics for ${\bf d}$, and in the absence of priors, the posterior distribution of  the model parameters $\theta$ is given by:
      \begin{equation}\label{eq:like}
        \chi^2\equiv-2\log p(\theta|{\bf d})=[{\bf d}-{\bf t}(\theta)]^T\,\hat{\sf C}^{-1}[{\bf d}-{\bf t}(\theta)],
      \end{equation}
      where ${\bf t}(\theta)$ is the theoretical prediction for ${\bf d}$, described in Section \ref{sec:theory}, and $\hat{\sf C}$ is the covariance matrix of our measurements.
      
      We build the covariance matrix $\hat{\sf C}$ as a block-diagonal matrix, where the first $N_\Xi\times N_\Xi$ block is given by the covariance matrix of the correlation function measurements (see Fig. \ref{fig:Cov_mat}). We assume the remaining $N_M+1$ elements (corresponding to the HI abundance in groups and the cosmic HI abundance) to be uncorrelated with the correlation function measurements, and  that their statistical uncertainties are also uncorrelated among themselves. These measurements are, however, correlated through some of their systematic uncertainties. In particular, the calibration of the absolute flux scale in ALFALFA dominates the systematic error budget in the measurement of $\Omega_{\rm HI}$ and $M_{\rm HI}^b$, and should affect all of these quantities in the same manner, rescaling them by an overall factor. In order to incorporate this correlation in our analysis we add, to the statistical covariance matrix described above, a systematic component that is fully correlated across the last $N_M+1$ measurements and with an amplitude $0.6\times10^{-4}$ in the $\Omega_{\rm HI}$-$\Omega_{\rm HI}$ component. Note that the measurement of the projected correlation function is immune to the effects of an overall rescaling factor, and therefore the corresponding part of the systematic contribution to the covariance matrix is fixed to 0.
      
      Finally, given the residual degeneracy between $M_{\rm min}$ and $M_0$ in our parametrization, we choose to show all our results in terms of $(\alpha,\log_{10}M_{\rm min},\Omega_{\rm HI})$ instead, but will also provide the corresponding best-fit value and uncertainty on $\log_{10}M_0$. We use broad top-hat priors for all parameters, with $c_{\rm HI,0}\in[0,100]$, $\alpha\in[0,2]$, $\log_{10}M_{\rm min}/h^{-1}M_\odot\in[8,13]$ and $\Omega_{\rm HI}\times10^4\in[0,20]$. In all cases we show constraints on $\log_{10}M_0,\,\log_{10}M_{\rm min}$ and $\alpha$ marginalized over the concentration parameter $c_{\rm HI,0}$.

      We sample the likelihood in Eq. \ref{eq:like} using the publicly available implementation of the Markov chain Monte Carlo algorithm {\tt emcee} \citep{2013PASP..125..306F}. The resulting constraints on the $\mhimh$ parameters are shown in Figure \ref{fig:triangle_fiducial} for different data combinations. We find compatible constraints from the clustering and groups data separately. Our marginalized constraints on the $\mhimh$ parameters are $\alpha=0.44\pm 0.08$, $\log_{10} M_{\rm min}/h^{-1}M_\odot=11.27^{+0.24}_{-0.30}$, $\log_{10} M_0/h^{-1}M_\odot=9.52^{+0.27}_{-0.33}$. The maximum-likelihood values are a good fit to the data in all cases, with a $\chi^2=13.7$ for 21 degrees of freedom for the full data vector. Although the clustering data is not able to jointly measure $\alpha$ and $\log_{10}M_{\rm min}$, and the groups data dominates the final uncertainties, clustering is still important in tightening the constraints (see e.g. the $\alpha$-$\log_{10}M_{\rm min}$ plane). In particular, we find that, within this model, the clustering measurements allow us to reduce the uncertainty on $\Omega_{\rm HI}$ with respect to the mass-function measurement of \cite{2018MNRAS.tmp..502J}, obtaining $\Omega_{\rm HI}=4.07^{+0.29}_{-0.26}\times10^{-4}$. 
      
      Figure \ref{fig:mhimh_summary} shows our best-fit $\mhimh$ relation (red solid line), together with its 1$\sigma$ uncertainty (shaded area) as well as our fiducial measurements of this relation on galaxy groups (blue points with error bars). The measurements from the DR7 group catalog are shown in blue. In order to jointly reproduce the measured HI content in high-mass halos as well as the measured total HI abundance, the model predicts a sharp drop in HI content below a halo mass $\log M_h/h^{-1}M_\odot\sim11.5$. The figure also shows, as black points, the $\mhimh$ relation measured in the IllustrisTNG-100 magneto-hydrodynamic simulation \citep{Paco_18} from a cosmological volume of $(75~h^{-1}{\rm Mpc})^3$. The errorbars represent the $1\sigma$ halo-to-halo variation on $\mhimh$. Although, overall, we find good agreement between our results and the simulation, for very small halo masses, the amplitude of $\mhimh$ differs significantly between our results and IllustrisTNG. This is however expected, given that the value of $\Omega_{\rm HI}$ in IllustrisTNG is $\simeq7.5\times10^{-4}$, i.e. roughly a factor of 2 larger than the ALFALFA measurement used here. Although our model predicts a larger low-mass cutoff than is found in simulations, existing data on halo masses below the range probed by the SDSS group catalog are not incompatible with this prediction. To illustrate this, Fig. \ref{fig:mhimh_summary} also shows the HI and halo masses measured for the Milky Way and M31 \citep{Draine_book, Braun_2009}.
      
      Finally, Figure \ref{fig:2pcf_best_fit} shows our measurement of the projected correlation function (blue points) together with the best-fit prediction and associated uncertainties (red line and shaded area) and the dark matter correlation function from HaloFit (black solid line) scaled by our our best-fit $b_{\rm HI}^2$ (see Section \ref{sec:discussion}). We also note that in contrast to \cite{Guo}, we are able to reproduce the measured HI clustering without involving assembly bias effects.

    \subsection{Impact of small scales} \label{ssec:results.smallscales}      
      On small scales, the halo-model prediction of the 2-point correlation function is dominated by the shape of the HI density profile. It is therefore important to evaluate whether our assumptions regarding the distribution of HI within each halo impacts our results on their overall HI content. 
      
\begin{figure*}
        \centering
        \includegraphics[width=0.49\textwidth]{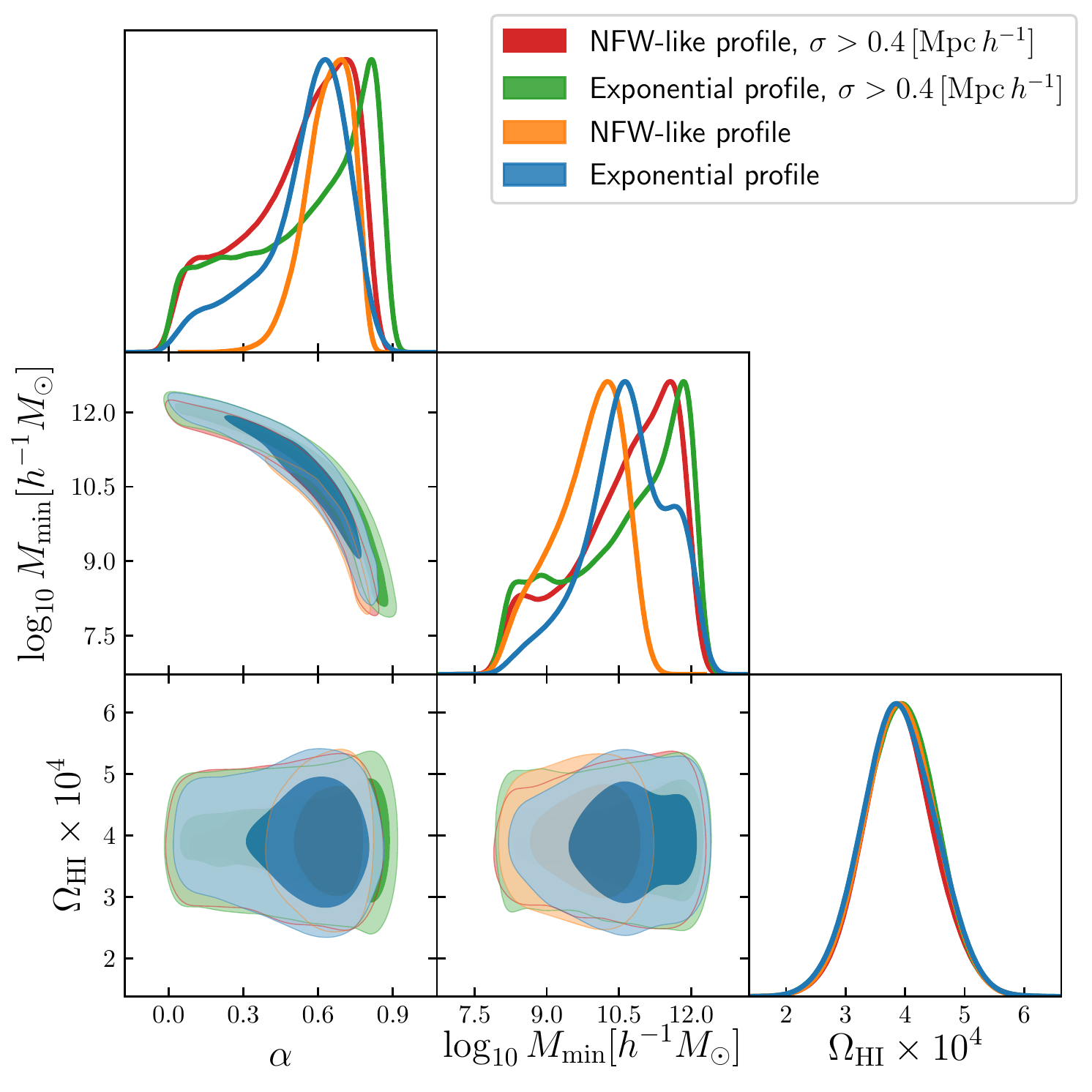}
        \includegraphics[width=0.49\textwidth]{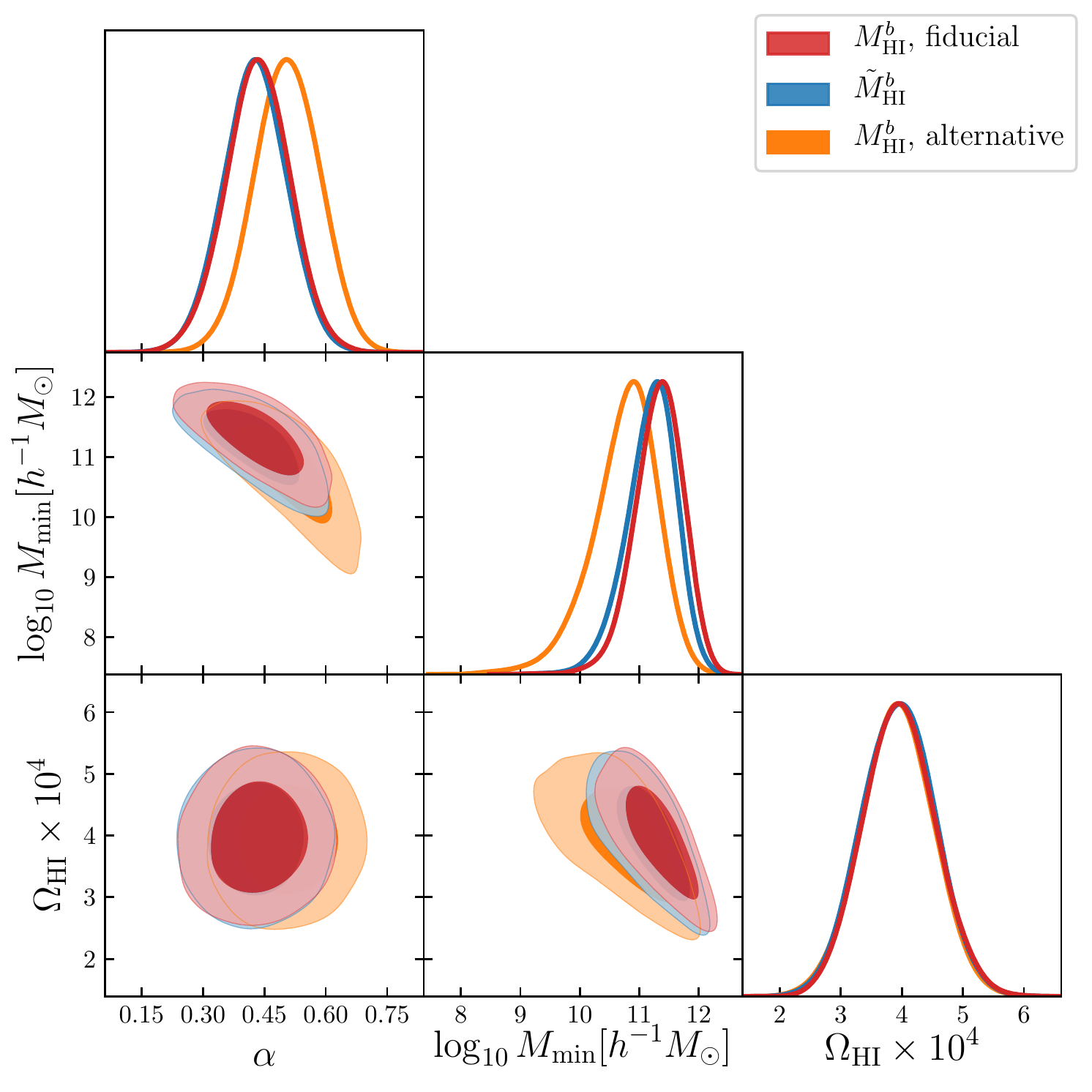}
        \caption{{\sl Left:} constraints on the $\mhimh$ relation derived from the clustering analysis under different scale cuts and choices of HI density profile. We derive our final scale cuts by demanding final constraints that do not depend on the choice of profile. {\sl Right:} constraints on the $\mhimh$ relation for different estimates of the HI mass in galaxy groups. Our fiducial measurements are shown in blue, while the red and light-orange contours show the results from the two alternative estimates described in Section \ref{ssec:method.hicont} (see also Fig. \ref{fig:MHIMF_comp}).}
        \label{fig:systematics}
      \end{figure*}

      The blue and light-orange contours in the top panel of Figure \ref{fig:systematics} show the constraints on the $\mhimh$ relation derived from the measurements of the projected correlation function for the exponential and altered NFW profiles described in Section \ref{sec:theory} respectively. Constraints are shown for the full range of scales ($\sigma\in(0.11,30.8)\,h^{-1}{\rm Mpc}$) and combined with the ALFALFA measurement of $\Omega_{\rm HI}$. The figure shows that the constraints on the $\mhimh$ parameters (particularly in terms of uncertainty) depend significantly on the model used to describe the distribution of HI within each halo. This is an undesirable feature, since we aim to constrain the global parameters of the $\mhi$-$M_h$ relation, given the large uncertainties in the actual shape of the HI density profile. On sufficiently large scales, in the 2-halo regime, this dependence should become negligible. We have verified this by removing all data points with $\sigma>0.43\,h^{-1}{\rm Mpc}$. These results are shown in Fig. \ref{fig:systematics} in green and red for the exponential and altered NFW profiles respectively. The dependence on the choice of profile, in terms of constraining power, vanishes in this regime. We thus use this restricted range of scales and the altered NFW profile for our fiducial analysis. Although the choice of profile in this regime is not relevant, we note that \citep{Paco_18} find that the altered NFW profile with an exponential cut-off on small scales is better able to fit measurements from hydrodynamical simulations.
   
    \subsection{Low-mass extrapolation} \label{ssec:results.lowmass}
      \begin{figure*}
        \centering
        \includegraphics[width=0.9\textwidth]{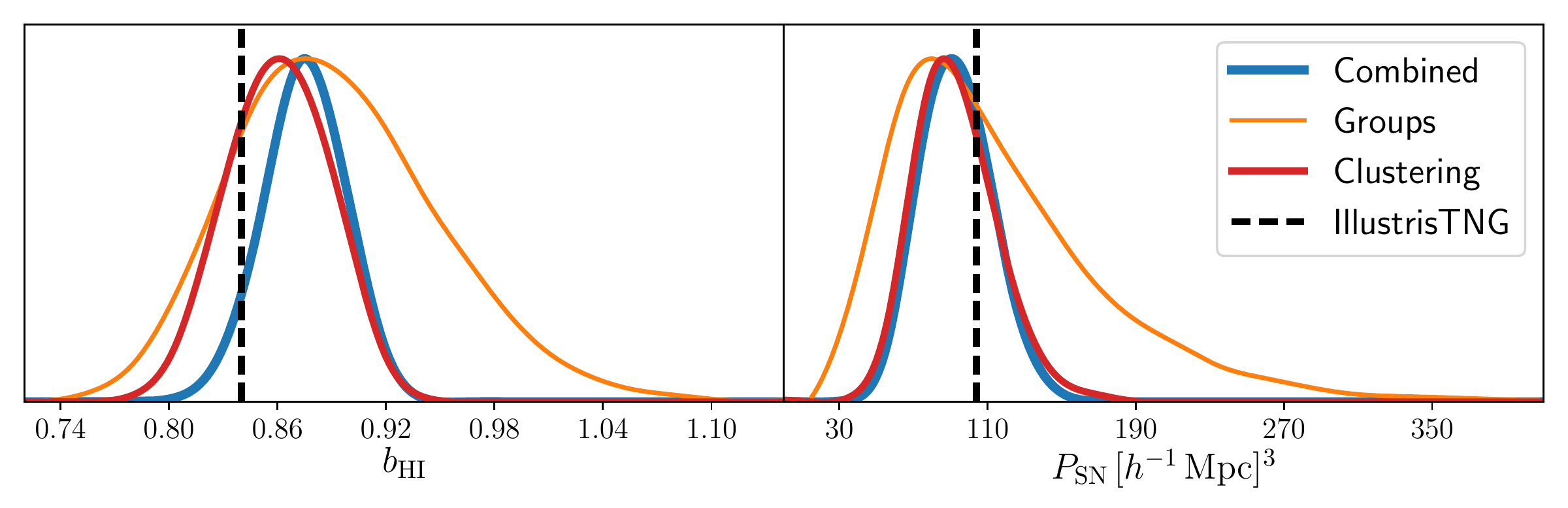}
        \caption{Posterior distributions for the large-scale HI bias ({\sl left}) and shot-noise power spectrum ({\sl right}) predicted from different combinations of our fiducial data vector: clustering$+\Omega_{\rm HI}$ (red), groups+$\Omega_{\rm HI}$ (light orange) and all data (blue). The vertical dashed line shows the result found in the IllustrisTNG simulation \protect\cite{Paco_18}.}
        \label{fig:bHI}
      \end{figure*}
      As described in Section \ref{ssec:method.hicont}, our measurement of the HI content of galaxy groups is based on extrapolating the HI mass functions measured in bins of halo mass beyond the detection threshold of ALFALFA. This is a legitimate approach as long as the range of masses covered by our sample constitute the main contribution to the total HI budget, in which case we only incurr in a small systematic effect when extrapolating the abundance of low-HI sources. We have shown that the mass deficit is generally below $\sim20\%$ of each individual HI mass measurement, and is always within the 1$\sigma$ uncertainties. The right panel of Fig. \ref{fig:systematics} shows the impact of this systematic on our final constraints on the parameters of the $\mhimh$ relation. The figure shows the constraints derived from our fiducial $\mhi$ measurements in red, as well as the contours corresponding to our two alternative estimates: summing over the 2DSWML mass function (blue) and re-scaling the global HI mass function (light orange). The constraints derived from both estimates are compatible, with a negligible shift in the best-fit $\log_{10}M_{\rm min}$. We therefore conclude that any residual systematics in the method used to measure the HI content as a function of halo mass in the group catalog is subdominant.
    
  \section{Discussion} \label{sec:discussion}
      We have placed constraints on the distribution of neutral hydrogen in dark matter halos as a function of halo mass. To do so we have used the HI-weighed clustering of 21cm sources detected by ALFALFA, as well as the abundance of those sources in halos identified in the galaxy group catalog compiled from the SDSS DR7 data. Our results show a power-law relation between $\mhi$ and $M_h$ at large halo masses with an exponent $\alpha=0.44\pm0.08$. This relation is exponentially suppressed on masses below $\log_{10}M_{\rm min}/h^{-1}M_\odot=11.27^{+0.24}_{-0.30}$. Although this suppression is not directly measurable in the data, given the mass range of the group catalog, it can be inferred indirectly by combining the group data with the total HI abundance measured by ALFALFA and our measurement of the 2-point correlation function.
      
      The constraints derived individually from our two datasets are compatible between themselves and with the combined constraints, and in all cases we find the model in Eq. \ref{eq:MHIM} to be a good fit to the data. It is worth emphasizing the fact that, although the clustering data is not able to break the degeneracy between $\alpha$ and $M_{\rm min}$, even when combined with the measurement of $\Omega_{\rm HI}$, it is vital to improve the constraints derived from the combination of the HI abundance in groups and $\Omega_{\rm HI}$. In fact we find that, within our model, clustering information is able to significantly reduce the final uncertainties on $\Omega_{\rm HI}$ compared with direct measurements of this quantity from the HI mass function. Furthermore, the clustering properties of the HI are arguably the most relevant piece of information for future 21cm intensity mapping studies, and this information is potentially better summarized by the projected correlation function data used here.
      
      Recently \cite{Paco_18} have aimed at characterizing the $\mhimh$ relation from state-of-the-art magneto-hydrodynamic simulations, and it is therefore relevant to explore the level of agreement between these simulated results and our data-driven constraints. In terms of the overall $\mhimh$ relation, this comparison is best summarized in Figure \ref{fig:MHIMF_comp}. We find that our results agree well with those of \cite{Paco_18} at $z=0$ for large halo masses ($M_{\rm h}\gtrsim10^{12.5}\,M_\odot/h$), and that our best-fit model as well, as the simulated data, are in good agreement with individual HI mass measurements. However, we observe that the $\mhimh$ relation derived from simulations departs significantly from our best-fit model on the low-mass end, predicting significantly higher HI masses. This disagreement is correlated with the higher value of $\Omega_{\rm HI}\sim7\times10^{-4}$ measured in IllustrisTNG, which is also the measurement that allows us to place constraints on the cutoff mass scale. The fact that the radiation from the sources is not accounted for in IllustrisTNG may explain the differences in the value of $\Omega_{\rm HI}$ and on the average HI mass inside small halos.
      
      For the purposes of predicting the clustering properties of HI in future 21cm experiments, two quantities are needed beyond $\Omega_{\rm HI}$: the large-scale HI bias $b_{\rm HI}$ and the shot-noise level $P_{\rm SN}$. Given our model for the $\mhimh$ relation, we can make predictions for these two quantities within the halo model ($b_{\rm HI}=F^1_1(k=0)$, $P_{\rm SN}=F^0_2(k=0)$, see Eq. \ref{eq:hmff}), which we can then directly compare with the values found by \cite{Paco_18}. The results of this comparison are shown in Figure \ref{fig:bHI}: our constraints on both quantities ($b_{\rm HI}=0.875\pm0.022$, $P_{\rm SN}=92^{+20}_{-18}\,[h^{-1}{\rm Mpc}]^3$) are in good agreement with the values predicted by IllustrisTNG at $z=0$. Although this result may seem at odds with the disagreement between data and simulation in terms of the total $\Omega_{\rm HI}$, this can be understood as due to the relatively higher contribution from larger-mass objects to these two quantities, for which our results agree with those of IllustrisTNG. It is also interesting to note that, even though the clustering data alone is not able to break the degeneracies between the $\mhimh$ parameters, they drive the constraints on both $b_{\rm HI}$ and $P_{\rm SN}$.
      
      Our measurement of the $\mhimh$ relation can be translated into a limiting circular velocity to host HI. Defining this as the circular velocity associated with a minimum halo mass such that 98\% of the cosmic HI is contained within heavier objects (see \cite{Paco_18}), we find $V_{\rm circ}=53^{+9}_{-11}\,{\rm km/s}$. This is in tension with the value found in \cite{Paco_18} ($V_{\rm circ}=34\,{\rm km/s}$), which is correlated with the higher cutoff halo mass measured in the data and shown in Figure \ref{fig:MHIMF_comp}.
      
      The results presented here are also interesting beyond future cosmological 21cm studies, as they provide insight into the distribution of neutral hydrogen across structures of different masses. Furthermore, our direct measurement of the $\mhimh$ relation is based on the characterization of the HI mass function for sources within galaxy groups, and have revealed hints about the relative dependence of the HI mass distribution on halo mass, with higher HI knee masses found on lower-mass halos. In general, the behaviour of the $\mhimh$ relation in the low-mass end ($M_h\lesssim10^{12}\,M_\odot/h$) is still somewhat uncertain, and its study will benefit in the future from higher-quality data and improved analysis methods. 
      
      We must also emphasize that the $\mhimh$ relation contains a huge amount of astrophysical information. In the high-mass end, the strength of processes such as AGN feedback, ram pressure and tidal stripping will leave its signature on the value of $\alpha$ \citep{Paco_16}, while on the low-mass end the presence of the UV background and the minimum mass to trigger self-shielding will determine the shape and amplitude of $\mhimh$. Our results can be used in combination with hydrodyanmic simulations or semi-analytic models \citep{Lagos_2014, Zoldan_2017} to improve our knowledge on the role of different astrophysical processes.

  \section*{Acknowledgments}
    We thank David Spergel for useful comments and discussions. AO and DA thank the Center for Computational Astrophysics, part of the Flatiron Institute of the Simons Foundation, for their hospitality. AO is supported by the INFN grant PD 51 INDARK. DA acknowledges support from the Beecroft trust and from the Science and Technology Facilities Council (STFC) through an Ernest Rutherford Fellowship, grant reference ST/P004474/1. The work of FVN is supported by the Simons Foundation. MGJ acknowledges support from the grant AYA2015-65973-C3-1-R (MINECO/FEDER, UE). We acknowledge the work of the entire ALFALFA collaboration team in observing, flagging, and extracting the catalog of galaxies used in this work. We also acknowledge the use of \texttt{CosmoMC/GetDist} \cite{Cosmomc_getdist}, \texttt{CAMB} \cite{CAMB}, \texttt{IPython} \cite{IPython}, \texttt{Matplotlib} \cite{Matplotlib} and \texttt{NumPy}/\texttt{SciPy} \cite{Numpy}.
    
\setlength{\bibhang}{2.0em}
\setlength\labelwidth{0.0em}
\bibliography{main}
%    \bibliographystyle{JHEP}
%    \bibliography{References.bib}

\end{document}